%% file: DriFi_v2.tex
\documentclass[conference, citesort]{IEEEtran}

\usepackage[numbers,sort&compress]{natbib}
\usepackage[utf8]{inputenc}
\usepackage{cleveref}
\usepackage{amsmath}
\usepackage[english]{babel}
\usepackage{booktabs}
\usepackage{hyperref}
\usepackage{url}
\usepackage{color}
\crefname{section}{§}{§§}
\Crefname{section}{§}{§§}
\usepackage{lipsum}

\newcommand\blfootnote[1]{%
  \begingroup
  \renewcommand\thefootnote{}\footnote{#1}%
  \addtocounter{footnote}{-1}%
  \endgroup
}
\ifCLASSINFOpdf
   \usepackage[pdftex]{graphicx}
   \graphicspath{{./figures/}}
\else
   \usepackage[dvips]{graphicx}
   \graphicspath{{../eps/}}
\fi

\begin{document}

\title{\bf \LARGE Mobile IMUs Reveal Driver's Identity From Vehicle Turns}

\newcommand*\samethanks[1][\value{footnote}]{\footnotemark[#1]}
\author{
\IEEEauthorblockN{Dongyao Chen\IEEEauthorrefmark{1},
Kyong-Tak Cho\IEEEauthorrefmark{1}, Kang G. Shin}
\IEEEauthorblockA{University of Michigan, Ann Arbor\\
\{chendy, ktcho, kgshin\}@umich.edu
}
}

\newcommand{\name}{\texttt{Dri-Fi}}
\newcommand{\appname}{\texttt{Dri-Fi}}

\maketitle
\thispagestyle{plain} 
\pagestyle{plain}
\input{body/abstract}
\blfootnote{*The authors equally contributed to this work.}
\input{body/introduction}
\input{body/threatModel}
\input{body/system}
\input{body/evaluation}
\input{body/relatedwork}
\input{body/discussion}

\input{body/conclusion}
\input{body/reference}

\IEEEpeerreviewmaketitle

\bibliographystyle{IEEEtran}

\end{document}

%% file: body/abstract.tex
\begin{abstract}
As vehicle maneuver data becomes abundant for assisted or autonomous driving, 
their implication of privacy invasion/leakage has become an increasing concern. 
In particular, the surface for fingerprinting a driver will expand significantly if the driver's 
identity can be linked with the data collected from his mobile or wearable devices 
which are widely deployed world-wide and have increasing sensing capabilities.

In line with this trend, this paper investigates a fast emerging driving data 
source that has driver's privacy implications. 
We first show that such privacy threats can be materialized via any mobile device with IMUs 
(e.g., gyroscope and accelerometer). We then present {\name} (\underline{Dri}ver \underline{Fi}ngerprint), 
a driving data analytic engine that can fingerprint the driver with vehicle turn(s). 
{\name} achieves this based on IMUs data taken only during the 
vehicle's turn(s). Such an approach expands the attack surface significantly 
compared to existing driver fingerprinting schemes.
From this data, \name\ extracts three new 
features --- acceleration along the end-of-turn axis, its deviation, and the deviation of the yaw rate --- 
and exploits them to identify the driver. Our extensive evaluation shows that an adversary equipped with 
\name\ can correctly fingerprint the driver within just {\em one turn} with 74.1\%, 83.5\%, and 90.8\% 
accuracy across 12, 8, and 5 drivers --- typical of an immediate family or close-friends circle --- respectively. 
Moreover,  with measurements 
on more than one turn, the adversary can achieve up to 95.3\%, 95.4\%, and 96.6\% 
accuracy across 12, 8, and 5 drivers, respectively.
\end{abstract}


%% file: body/introduction.tex
\section{Introduction}
\label{sec:intro}
As data of vehicle maneuver becomes abundant for assisted or autonomous driving, 
their implication of privacy invasion/leakage has become an increasing concern. 
To prevent potential privacy violations, the U.S. Congress has enacted a law for enforcing driving 
data privacy in Dec. 2015~\cite{driver_law}. In particular, the law forbids disclosure of
personally identifiable information of the owner or the lessee of the vehicle. 
In Dec. 2016, NHTSA also enforced the protection of any data that can be ``reasonably linkable'' 
to driver identification~\cite{nhtsa_v2v_privacy}.

Despite these legislations, researchers have demonstrated that driver's 
privacy can indeed be breached by accessing in-vehicle data through an On-Board 
Diagnostics (OBD-II) dongle.
For example, 
the authors of \cite{yoshi2016,single_turn2016} showed that 
the driver's identity can be revealed by analyzing the vehicle's 
Controller Area Network (CAN) data collected through the OBD-II port.
Although this could be a severe privacy threat, its practicability/feasibility
has been questioned for two reasons. First, due to security concerns, 
car manufacturers are beginning to restrict the OBD-II port access, i.e., allowing 
its access only during diagnostics (while the vehicle is parked)~\cite{obdblock}. 
Second, even with OBD-II access, existing driver-fingerprinting schemes 
require a time-consuming task of reverse engineering in-vehicle 
data~\cite{cids, busoff}. All of these together make it very difficult to 
invade the driver's privacy via the OBD-II port.

Due to the nature of in-vehicle data being obscure and difficult to access 
(e.g., physical access to one's car), researchers/developers increasingly 
use IMUs --- available on various devices such as smartphones, OEM-authorized 
OBD-II dongles, and wearables --- as an alternative source of driving data for 
enhancing driving experience and safety.
This use of IMUs in the automotive ecosystem has led to the development of various 
``beneficial'' (c.f. malicious) applications such as
driving-assistance systems~\cite{truemotion}, adjustable auto insurance~\cite{ubi_wiki},  
and fuel-efficient navigations.

Collection and exploitation of IMU data also create concerns of breaching drivers' privacy. 
In particular, data-collection entities might be able to infer the driver's identity from the collected IMU data, 
leading to an incontrovertible breach of the driver's privacy. 
This paper focuses on the driver's identity privacy, and hence questions 
``Would existing schemes on mobile devices breach the driver's privacy?
	Can an adversary with access to only IMU data achieve it?''


On one hand, researchers have shown that one's privacy can be breached if 
his/her device is identified/tracked via stealthy identifiers available on the device.
For example, by leveraging 
the imperfection of IMU components~\cite{accel_NDSS2014,boneh_fingerprint} or 
non-cookie web tracking techniques (e.g., supercookies~\cite{web_tracking}) on 
a mobile device, an adversary can identify the device and/or its user.
On the other hand, instead of identifying the device itself (and hence its owner), 
other existing schemes attempt to identify the user 
through his/her behavior or interaction with the device 
(e.g., touch screen behavior~\cite{silent_sense}, DNS traffic pattern~\cite{dns_tracking}). 
Although these existing schemes indeed breach privacy of the {\em device 
owner/user}, they do not necessarily breach the {\em actual driver's} privacy. 
For example, suppose driving data was collected from a smartphone while 
its owner was in a car as a passenger. 
In such a case, the collected data did not originate from the actual driver's 
device, and hence will not help identify the driver.
Similarly, existing schemes cannot identify the driver when someone simply takes his phone 
and then goes on for a drive.
Meanwhile, an interesting but yet unanswered question is ``if an 
adversary reads and analyzes the IMU data in more depth, would the consequences 
be different?'' 
Behind the paradigm shift of how devices (equipped with IMUs) are being 
used/integrated in contemporary automotive ecosystems (e.g., 
vehicle authentication via smartphones, event data recording via IMUs), there 
could exist many uncovered scenarios where the driver's privacy could be 
unintentionally breached.

In this paper, we propose a new driver fingerprinting scheme called
{\name} (\underline{Dri}ver \underline{Fi}ngerprint), which can be used by an adversary 
to stealthily infer the driver's identity based on the readings of zero-permission mobile IMUs,
i.e., gyroscope, accelerometer, and magnetometer.
By developing and using {\name}, we focus on the risk of breaching the driver's privacy 
based on driving data that is far more easily obtainable and accessible 
than in-vehicle data, i.e., IMU data. 

The key challenge for {\name} to fingerprint drivers is that it has much less 
information available than the existing schemes that use in-vehicle 
data~\cite{yoshi2016, single_turn2016}; 
all \name\ has is access to IMU sensor readings.
\name\ overcomes this challenge by constructing a driving behavior profile 
based on the drivers' turning maneuver --- a common yet representative maneuver.
There are two reasons for using vehicle ``turns'' to represent the driver's 
behavior: 1) turns are 
behavior-rich actions that reflect how the driver accelerates/decelerates, and at the same time, 
how s/he steers, and 2) turns are less likely to be affected by traffic 
conditions than other maneuvers. 
For example, deceleration can be affected greatly by the frontal car whereas turns are not. 
So, once \name\ detects a {\em turn}, it derives three new features: 
1) the {\em acceleration along the end-of-turn axis}, which is defined as the axis 
orthogonal to the vehicle's direction when the turn started, 
2) the {\em deviation in the first feature}, and 
3) the {\em deviation in the yaw rate}.
These features function as the cornerstone of \name's driver fingerprinting as 
they reflect the driver's unique behavior --- they are only affected by how the 
driver actually turns the steering wheel or how s/he presses the 
acceleration/brake pedal while making a turn.
Our extensive experimental evaluation will later show that these 
features vary only with drivers, but {\em not} with car models or trip routes. 
Once these three features are derived for a detected turn, \name\ computes various percentiles and 
their autocorrelations which are then used to construct the feature vector for 
machine classifiers, such as Naive Bayes, SVM, and Random Forest.  
\name\ can, therefore, fingerprint the driver with a high probability, even when s/he makes just one left/right turn. 
As more turns are made by the driver within a trip, \name\ exploits such accumulated information 
for driver fingerprinting to reduce false positives/negatives, enhancing its accuracy.
In addition to the driver fingerprinting, we also discuss how an adversary may
utilize its IMU measurements to construct a well-formulated training dataset 
from scratch, which is essential for the underlying machine classifiers. 
Note that all existing studies (had to) assume the training dataset, which 
has the correct labels of all targeted drivers, was {\em given} to the adversary, which may not hold in practice.

We evaluated \name\ extensively by collecting IMU data from a smartphone while 
12 different drivers (9 males, 3 females) were driving either around the campus or in an urban/rural area. 
Our results have shown that by using \name, the driver can be  
identified within one left/right turn, with accuracies of 74.1\%, 83.5\%, and 
90.8\% across 
12, 8, and 5 drivers\footnote{The selection of this number of drivers is to reflect real-life scenarios, 
covering immediate family members, close friends and colleagues, etc.} respectively. 
 \name's achievement of high accuracy {\em with just one turn} implies a severe 
 driver privacy invasion. 
Also, \name's performance is in sharp contrast to existing studies that require 
several minutes of measurements from tens of in-car sensors. 
As the driver made 8 turns, \name\ 
was able to achieve accuracies of 95.3\%, 95.4\%, and 96.6\% 
across 12, 8, and 5 drivers, respectively.

This paper makes the following main contributions:
\begin{itemize}
\item[1.] Discovery of three new features which are shown to
	be distinct between different drivers and independent of vehicles and trip 
	routes (Sec.~\ref{sec:system});
\item[2.] Development of \name, which extracts the new 
	features when the driver makes a left/right turn, and thus achieves 
	driver fingerprinting (as soon as the driver makes a turn) with high 
	accuracy (Sec.~\ref{sec:system});
\item[3.] Implementation and extensive evaluation of \name\ using commodity 
smartphones (Sec.~\ref{sec:evaluation}).
\end{itemize}
%

%% file: body/threatModel.tex
\section{Motivation, Adversary Model, and Goal}\label{sec:threatModel}
Although the abundance of driving-related data has brought 
various benefits to our lives, their availability can also breach the driver's privacy. 
\subsection{Why Not Fingerprinting Drivers with In-car Data?}
An adversary with access to sensors on an in-vehicle network, such as the Controller Area Network (CAN), 
can fingerprint the driver \cite{yoshi2016, single_turn2016, 
driver_inertial_sensor, driverID_gmm2007, knowmaster}. 
(Such schemes will be detailed in Sec.~\ref{sec:relatedwork}.)
Despite the rich and low-noise in-car data for the adversary to fingerprint drivers, 
s/he must meet the following two minimum requirements to acquire the data, 
which are assumed to have been met in all existing studies.

{\bf Access to In-car Data.}
To read and extract values of sensors on an in-vehicle network, the 
adversary must have access to the sensors data. 
To gain such an access, s/he may either 1) remotely compromise an Electronic 
Control Unit (ECU), 
or 2) have a compromised  OBD-II dongle plugged in the victim's vehicle in order to read
in-car data.\footnote{Note that drivers may try to lower their auto-insurance 
rates by plugging in OBD-II dongles provided by the insurance companies, or 
use them for 24/7 monitoring of the health of their cars. It has been shown that these 
dongles can also be compromised by adversaries~\cite{woot_tele}.}
For the first case, however, depending on the ECU that the adversary 
compromised, s/he may not be able to read {\em all} sensors data of 
interest, mainly because the ECUs which produce those data may reside 
in different in-vehicle networks (e.g., some on high-speed CAN and others on low-speed CAN).
For the second case, the adversary has indeed control of a plugged-in and compromised OBD-II dongle, 
and therefore, in contrast to a compromised ECU, is likely to have access to all sensors data of interest 
(as shown in~\cite{yoshi2016,knowmaster}). However, for security reasons, car 
manufacturers are increasingly blocking/restricting in-car data access through 
the OBD-II port except when the vehicle is parked~\cite{obdblock}.
Thus, the adversary will less likely be able to access in-car data. 

{\bf Reverse-engineering Messages.}
Even when the adversary has access to in-vehicle network messages,  
s/he must still (i) understand where and in which message the sensor data (of interest) is contained, 
and  (ii) translate them into actual sensor values (e.g., transformation coefficients 
for addition/multiplication of raw sensor data~\cite{yoshi2016}). 
In-vehicle network messages are encoded by the vehicle manufacturers and the ``decoding book,'' 
which allows one to translate the raw data is proprietary to them. Therefore, unless the adversary has 
access to such a translator, s/he would have to reverse-engineer the messages, 
which is often painstaking and incomplete.

Although the adversary may have abundant resources to fingerprint
the driver, meeting the above two requirements may be difficult or even not possible.

\subsection{Adversary Model}\label{sec:adver_model}
Due to the difficulty and (even) impracticality of an adversary fingerprinting 
the driver via in-vehicle (CAN-bus) data, we consider the following adversary who might 
fingerprint the driver without the difficulties of state-of-the-art solutions. 
In particular, we consider the adversary with a data-collection entity that 
aims to fingerprint the driver based on zero-permission {\em mobile IMU data}. 
We assume that the adversary has access to the target's mobile IMU data 
while s/he was driving. As mobile IMUs are available in various commodity 
mobile/wearable devices such as smartphones, watches, and even in OBD-II dongles, the 
adversary can compromise one of them (belonging to the target), and
obtain the required IMU data for driver fingerprinting.
This means that the adversary would have a much larger attack surface than 
existing driver fingerprinting schemes.
One example of such an adversary would be a smartphone malware programmer who 
builds an app to stealthily collect the target's IMU data.
Another example could be a car insurance company that might reveal information 
other than what was initially agreed on via the collected/stored IMU data 
available on its OBD-II dongles.

\subsection{Motivating Scenarios}\label{subsec:motivating_cases}
Integrating mobile IMU sensors with the automotive ecosystem can, on one hand, 
lead to development of numerous beneficial apps. 
On the other hand, it may violate the driver's privacy. 
In what follows, we state three double-edged-sword scenarios, which at 
first glance seem beneficial for our daily driving experience but could lead to a severe privacy violation; 
in fact, more severe than what has already been studied/uncovered.

\subsubsection{Vehicle Authentication}
To enable a more convenient car-sharing experience, car companies, such as 
Volvo~\cite{volvo_keyless} and Tesla~\cite{tesla_keyless}, started to let car owners unlock and start their 
cars (e.g. new Tesla model 3) by using their smartphone apps, thus replacing a key fob with a smartphone. 
By installing this authorized app, the car owner first designates eligible drivers as a whitelist. 
All allowed drivers can then unlock and start the car with authentication through the 
Bluetooth link between the car and their smartphones.

\noindent\textbf{Privacy violation case.}
Alice owns a car with this functionality. Her husband Bob's driver's license was 
suspended. So, Alice is unable to register him as a driver in 
the whitelist, due to a background check conducted by the car company. 
One day, Alice asks Bob to drive the car for some reasons. To 
evade the driver authentication, Alice temporarily gives Bob her phone 
to drive the car. However, if the car company's app had stored IMU data and 
thus had the driving profiles of all whitelisted drivers, with the capability of 
identifying the driver from IMU data, the car company can determine that the 
current driving pattern (which is Bob's) does not match with any of the whitelisted.
This becomes a definite privacy violation if the car company had 
initially stated/claimed that all the IMU data (while driving) reveals
how the car moves, not who actually drives it. However, the driver's identity 
can be found via in-depth analysis.
%

\subsubsection{Named Driver Exclusion}
Many states in the U.S. permit ``named driver exclusion'' to allow auto insurance 
buyers to reduce their premium~\cite{insurance_exclusion, progressive_exclude}. 
Under this plan, the insurance company will not accept any excuses for 
allowing the excluded person to drive. 
Therefore, Department of Motor Vehicles (DMV) specifically warns all drivers of the fact that, 
to avoid driving without any insurance coverage, 
the excluded individuals should not drive the insuree's car~\cite{dmv_alert}. 

\noindent\textbf{Privacy violation case.} Suppose Bob's wife, Alice, is a legitimate 
driver. However, to reduce the cost of their family insurance plan, Bob 
excludes Alice from the plan. Bob's smartphone has installed the insurance 
company's app, which not only manages his insurance account but also 
keeps record of the driving IMU data as an Event Data Recorder 
(EDR).\footnote{EDR is used for getting a detailed picture of the seconds right 
before and after a crash.}
One night, Bob was in a bad physical condition and hence asked 
Alice to drive him home. 
Unfortunately, they ran into an incident. 
At the court, the insurance company defended 
itself by showing the driving IMU data --- measured during that night when the 
accident occurred --- matched Alice's, not Bob's, driving profile. Thus, the 
company refused to reimburse Bob and won the lawsuit.
Note that the initial purpose of EDR functionality on the app was not for 
driver fingerprinting but for recording events, an undetermined privacy violation.

\subsubsection{Utilization of IMU Data}
Unlike conventional OBD-II dongles (designed for diagnostics), 
car manufacturers are designing and developing a new type of dongle, which 
does {\em not} provide users with raw CAN data but provides them 
in a ``translated'' format (e.g., JSON format). 
Ford OpenXC~\cite{ford_openxc} and Intel-based OBD-II dongles~\cite{intel_obd} 
are examples of such a design. 
This way, the car OEMs' plugged-in dongle reads and translates metrics 
from the car's internal network and provides them to the user {\em without} 
revealing proprietary information. 
Thus, while providing the {\em necessary} information to the users, car OEMs can let 
them install vehicle-aware apps which have better interfaces based on a context  
that can minimize distraction while driving~\cite{ford_openxc}.

\noindent\textbf{Privacy violation case.}
Alice has the car OEM's dongle, which provides her the translated CAN data, 
plugged in her car so that she can gain more insight into her car operation.
Due to a security breach on the dongle, suppose Mallory has access to the 
data being read from the dongle, but only in a translated format. 
Note that even with access to raw CAN data, 
Mallory would still need to reverse engineer the messages; we are relaxing the 
technical requirements for Mallory. He may fail because the translated data that Mallory has access 
might not contain the required information for in-car-data-based driver fingerprinting. 
Note that the most significant feature used for driver fingerprinting in \cite{yoshi2016} was the 
brake pedal position, which unfortunately is not provided by the Ford OpenXC~\cite{ford_openxc}.
However, since those dongles are always equipped with IMUs for data 
calibration, Mallory uses his malware 
to read the IMUs instead, and thus attempts to identify the driver.
This implies that Mallory might not even need to access the 
translated data at all, thus lowering the technical barrier for the adversary.
Through security by obscurity, the translation of data itself might provide 
some sort of privacy. However, the IMUs installed on those dongles, 
designed for calibration, might ironically threaten the driver's privacy.

\subsection{Our Goal}\label{sec:problem}
To breach the driver's privacy, the adversary needs an efficient way of 
fingerprinting the driver solely based on IMU data.
Researchers have already demonstrated the feasibility of an adversary breaching 
the driver's privacy by fingerprinting him/her with in-car data. We refer to such 
an adversary as a {\em high-resource adversary} due to his/her access to the rich and 
low-noise in-car data. However, we still do not know if a 
{\em low-resource adversary}, with access to only the target's IMU data, can fingerprint the driver; 
it may even be infeasible due to his/her insufficient resource(s).
Therefore, our goal is to shed light on an unexplored but important question:
%
``Within a short duration, can a low-resource adversary fingerprint 
the driver, i.e., having access to only IMUs?"

%% file: body/system.tex
\section{System Design}\label{sec:system}
\input{body/system01}

\input{body/system02}

%% file: body/system01.tex
We propose \name, which acquires IMU sensor measurements from the victim/target
driver and exploits them for identifying the driver.

\subsection{Overview of {\name}}
\label{sec:overview}
\begin{figure}
	\centering
	\includegraphics[width=\linewidth]{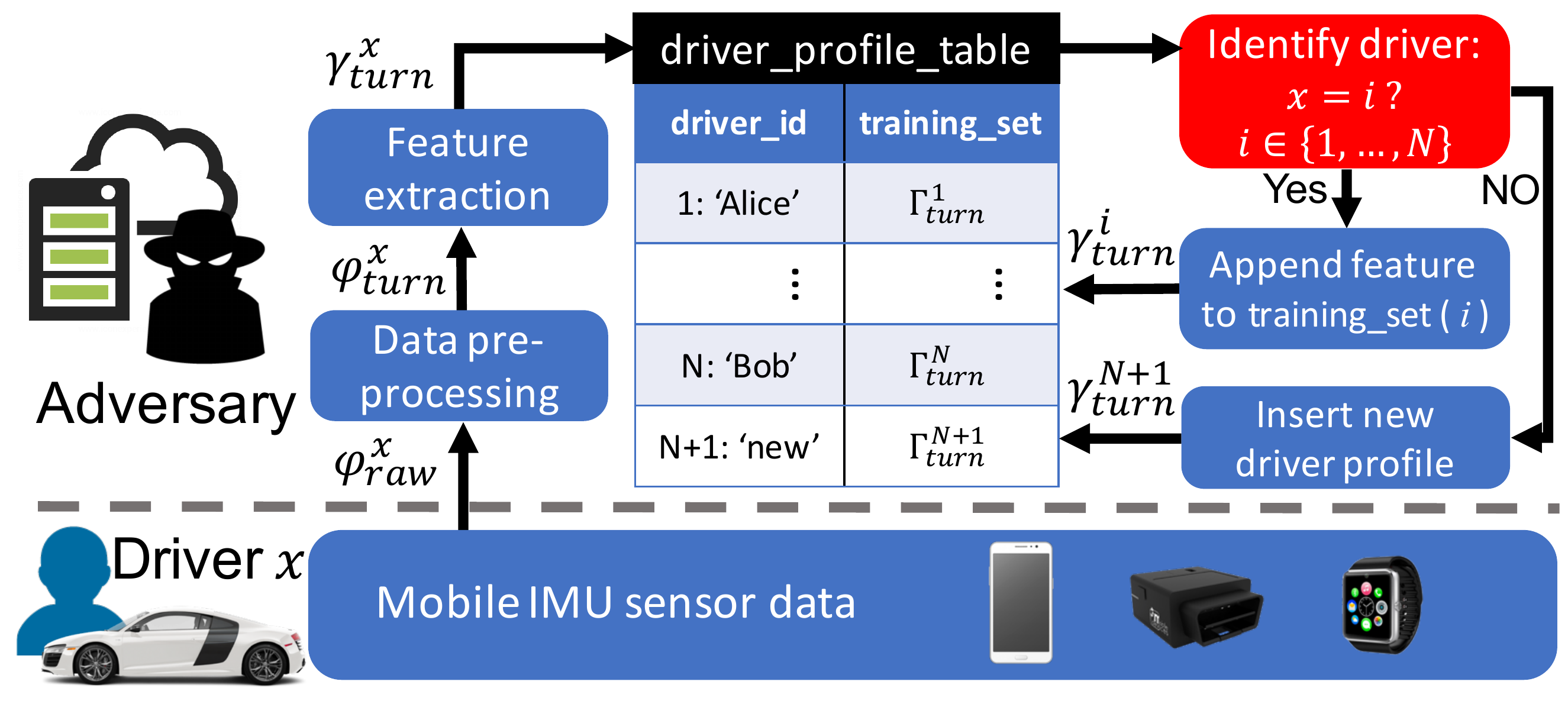}
	\vspace{-2em}
	\caption{System overview of {\name}.}
	\label{fig:overview}
	\vspace{-1em}
\end{figure}

Fig.~\ref{fig:overview} presents a high-level overview of {\name} as a 
4-step process where the adversary acquires the required raw IMU sensor 
data $\varphi^{x}_{raw}$ for fingerprinting driver $x$.
The data can be acquired either at the end of, or periodically during a trip.
First, \name\ pre-processes $\varphi^{x}_{raw}$ to remove noises and extracts the sensor measurements 
only while $x$ was making a (left/right) turn (Sections~\ref{subsec:data_collection}--\ref{subsec:turnextract}). 
So, it obtains $\varphi^{x}_{turn}$.
Next, based on the thus-obtained $\varphi^{x}_{turn}$, \name\ constructs a set 
of features, i.e., a feature vector $\gamma^{x}_{turn}$.
Then, based on a Gaussian Mixture Model (GMM) (to be detailed
in Section~\ref{subsec:trainset}), the adversary first verifies 
whether $\gamma^{x}_{turn}$ closely matches any of his/her previously obtained 
training data $\Gamma^{n}_{turn}$ in the driver\textunderscore 
profile\textunderscore table, where $n\in \{1,\ldots,N\}$ 
and $N$ is the number of drivers the adversary had learned about.
If there is a close match, s/he exploits $\gamma^{x}_{turn}$ as an input for 
machine classifiers with the training set as driver\textunderscore 
profile\textunderscore table.
As a result, s/he identifies the driver $x$ to be $i \in \{1,\ldots,N\}$
(Sections~\ref{subsec:man_fingerprint}--\ref{subsec:trip_fingerprint}).
Finally, s/he appends $\gamma^{x}_{turn}$ to his/her 
driver\textunderscore profile\textunderscore table with label $i$.
Meanwhile, based on GMM, if $\gamma^{x}_{turn}$ does not closely match any 
of those in the driver\textunderscore 
profile\textunderscore table, s/he constructs 
a new driver training dataset for driver $x$. 

Note that the adversary needs to have driver\textunderscore profile\textunderscore table constructed 
before identifying driver $x$.
Due to this requirement, existing driver fingerprinting schemes with CAN data assume that 
the adversary already has a complete training set for all the targeted drivers.
However, this assumption is difficult to be met in practice. In the following 
sections, to focus on the design and discussion of \name, for now, we assume 
that the adversary does have the complete training set; the same assumption as 
in previous studies. 
We will present in Section~\ref{subsec:trainset} how such an assumption 
can be relaxed by using our context-based approach for constructing the training set from scratch.

\subsection{Data Collection and Pre-processing}
\label{subsec:data_collection}
{\name} continuously collects the raw IMU data $\varphi_{raw} = \{gyro_{raw}, 
acc_{raw}, mag_{raw}\}$ --- raw readings from the gyroscope, accelerometer,
and magnetometer, respectively --- throughout a trip.
To accommodate different postures of the mobile device, which \name\ utilizes for data
collection, inside the car, {\name} aligns the coordinate of the IMU readings using 
the magnetometer~\cite{vsense}.\footnote{If \name\ were to use sensor measurements from 
an OBD-II dongle, since that device's posture would not change, \name\ need not align
the coordinates, thus not requiring the use of the magnetometer.}
Specifically, {\name} always aligns the device's coordinate with the
geo-frame/earth coordinate so as to maintain the consistency of analysis.
This allows the data which \name\ uses for driver fingerprinting to be not
affected by the device postures, i.e., it works under various placements/circumstances.

Once the coordinate-aligned data of the gyroscope and accelerometer sensors
have been collected, {\name} smooths and trims them to prepare for further analyses.
If the device which \name\ uses is a smartphone, its handling by the user may 
cause
high-power noises on the gyroscope and accelerometer sensors.
Abnormal road conditions (e.g., potholes) may also incur a similar level of noise. 
Therefore, \name\ first removes those noises by filtering out abnormal spikes in the data. 
\name\ then smooths each IMU sensor (gyroscope and accelerometer) data stream by 
using a low-pass filter to remove high-frequency noises.

\subsection{Extraction of Left/Right Turns}
\label{subsec:turnextract}
\name\ trims the smoothed data further by retaining the IMU measurements
acquired only during a left/right turn, $\varphi_{turn} = \{gyro_{turn}, acc_{turn}, mag_{turn}\}$, 
i.e., smoothed IMU data of only left/right turning maneuvers. 
In other words, measurements taken when the driver constantly drove on a straight road, 
or when the car stopped to wait for traffic lights or stop signs are all discarded.
Among the various maneuvers (e.g., turns, lane changes, acceleration/deceleration),
the reason for \name's focus on data from left and right turns is that
the vehicle/driver's actions/maneuvers for making turns are much less likely
affected by the car in front (i.e., traffic condition) than others.
For example, deceleration of a vehicle would depend on the car in front, whereas
left/right turns are less likely to depend on it.

In order to extract data only related to left/right turns, among the three IMU
sensors --- gyroscope, accelerometer, and (perhaps) magnetometer --- \name\ 
uses the (coordinate-aligned)
gyroscope's yaw rate reading as it reflects the vehicle's angular velocity around its vertical axis, 
i.e., the vehicle's rotational inertia.\footnote{Accelerometer readings after the coordinate alignment would only 
show the changes in the longitudinal/lateral acceleration in reference to the vehicle's heading direction.}
Note, however, that a non-zero value from the gyroscope does not necessarily
represent a left/right turn, since there exist other (similar) maneuvers such
as lane changes and U-turns which incur similar results~\cite{vsense}.
Hence, based on the gyroscope readings, \name\ extracts data of only left/right
turns in the following two steps: it
\begin{itemize}
\item [{\bf S1.}] Recognizes whether or not a steering maneuver --- which we refer to as
{\em maneuvers} (left/right turns, lane changes, U-turn, etc.) that suddenly
change the vehicle's heading direction significantly --- was made;
\item [{\bf S2.}] Determines whether the steering maneuver was a left/right turn and,
if so, extracts sensor readings acquired during that turn.
\end{itemize}

\begin{figure}
	\centering
	\includegraphics[width=0.9\linewidth]{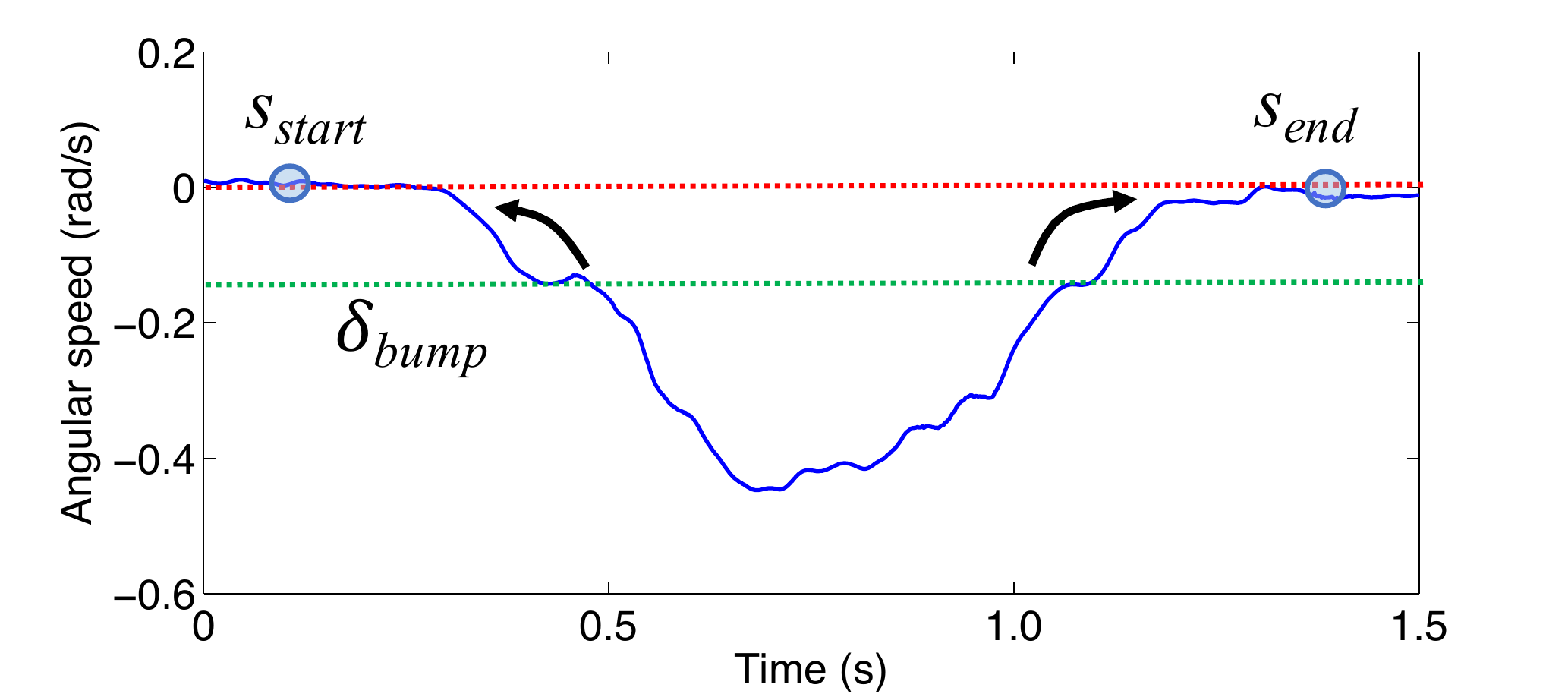}
	\vspace{-1em}
	\caption{Left turn extraction from smoothed gyroscope readings.}
	\vspace{-1em}
	\label{fig:sensor_turn}
\end{figure}

\textbf{S1. Recognizing steering maneuvers.}
\name\ recognizes the occurrence of a steering maneuver when the yaw rate
readings from the gyroscope form a ``bump-shape''.
When a car changes its direction by making a left turn, as shown in
Fig.~\ref{fig:sensor_turn}, the yaw rate reading from the gyroscope first
decreases, reaches its minimum peak, and finally rises back to approximately
0 rad/s when the left turn is completed.
For a right turn, everything would be the opposite to a left turn; increase, reach the
maximum, and decrease. Depending on how the coordinates are aligned, a
negative bump may reflect a right turn, not a left turn. However, in this paper,
we consider the yaw rate to increase when rotated clock-wise.
Based on this observation, \name\ determines that a steering maneuver has
occurred if the absolute yaw rate exceeds a certain threshold,
$\delta_{bump}$, which is empirically set to 0.15\, $rad/s$. 
Note that without the
threshold ($\delta_{bump}$), even a small movement of the steering wheel would
cause \name\ to mis-detect a steering maneuver.
Thus, \name\ marks the start time/point of that steering maneuver as
$s_{start}$ when the absolute yaw rate, $|Y|$, exceeded
$\delta_{bump}$ for the first time. Also, \name\ marks the end point, $s_{end}$, as when $|Y|$
first drops back below $\delta_{bump}$.
Since the steering would in fact have started a bit before
$s_{start}$ and ended a bit later than $s_{end}$, where $|Y|\approx0$ as
shown in Fig.~\ref{fig:sensor_turn}, \name\ moves points $s_{start}$ and
$s_{end}$ backwards and forwards, respectively, until $|Y|\approx 0$.
As a result, \name\ interprets a steering maneuver to have made at a time
within $s = [s_{start}, s_{end}]$.

\begin{figure}
	\centering
	\includegraphics[width=0.95\linewidth]{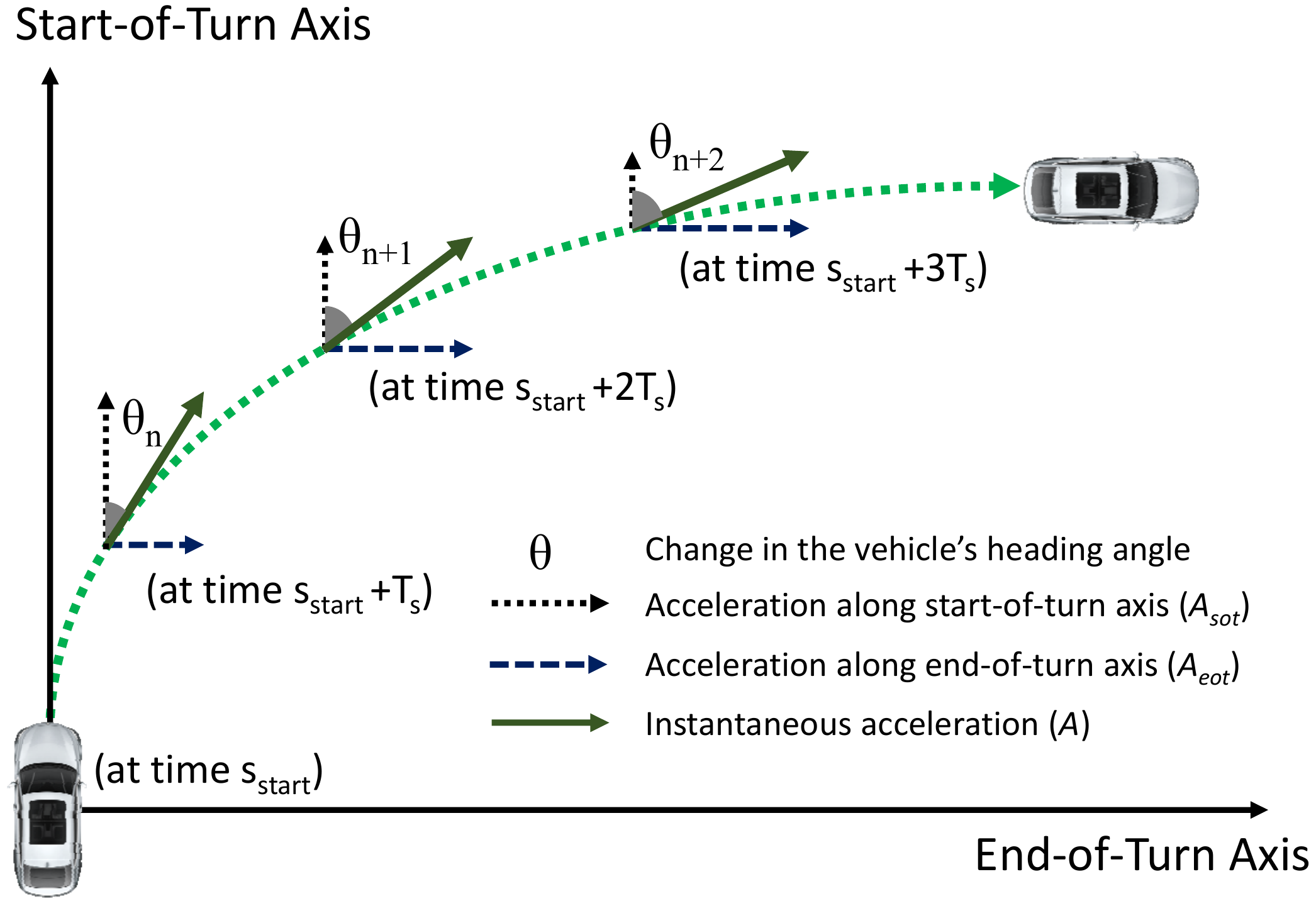}
	\vspace{-1em}
	\caption{Changes in the vehicle's accelerations and heading angle.}
	\vspace{-1em}
	\label{fig:headingaxis}
\end{figure}
\textbf{S2. Filtering left/right turns.}
The steering maneuver extracted in S1 may be comprised of not only left/right
turns but also lane changes or U-turns, since those maneuvers yield similar
bump-shaped yaw rate readings.
In order to extract only left/right turns, as in \cite{vsense}, \name\ derives
the change in the vehicle's heading angle, which
is defined as the difference in the vehicle's heading angle between the start
and the end of a steering maneuver.
Fig.~\ref{fig:headingaxis} shows an example vehicle trajectory during a right
turn where three IMU sensor readings were acquired at times $t=s_{start}+\{T_s,
2T_s, 3T_s\}$, i.e., sampled with frequency of $1/T_s$.
As in step S1, let $t=s_{start}$ be the time when the vehicle was detected to
have started the turn. Since the yaw rate readings from the gyroscope represent
the vehicle's angular velocity around the vertical (Z) axis, the change in the
vehicle's heading angle after time $nT_s$ has elapsed since $s_{start}$,
$\theta[nT_s]$, can be approximated as
\vskip -7pt
\begin{equation}
\theta[nT_s]\approx\theta[(n-1)T_s] + {Y_{n} T_s}=\sum\limits_{k=1}^{n} {Y_k
T_s},
\end{equation}
\noindent where $Y_n$ denotes the $n$-th yaw rate reading since $t=s_{start}$.
Therefore, at the end of making a right turn, the eventual change in the vehicle's heading angle,
$\theta_{final}=\theta[s_{end}-s_{start}]$ would be approximately 90$^\circ$
whereas at the end of a left turn it would be -90$^\circ$.
This change in the vehicle's heading angle is a good indicator for determining
whether the vehicle has made a left/right turn, since for lane changes,
$\theta_{final}\approx0^\circ$ whereas for U-turns, $\theta_{final}\approx180^\circ$.
Thus, \name\ calculates the $\theta_{final}$ of a detected steering maneuver
(made during $s_{start}\sim s_{end}$), and only retains it such that
$70^\circ \leq |\theta_{final}| \leq 110^\circ$, i.e., approximately $\pm$90$^\circ$.
Note that since left/right turns usually take a short period
of time ($<$3 seconds), drifting in the gyroscope during a turn~\cite{inferroute}
does not affect \name's performance.

\begin{figure}[t]
	\begin{minipage}[t]{0.19\textwidth}
		\centerline{\includegraphics[width=4.0cm]{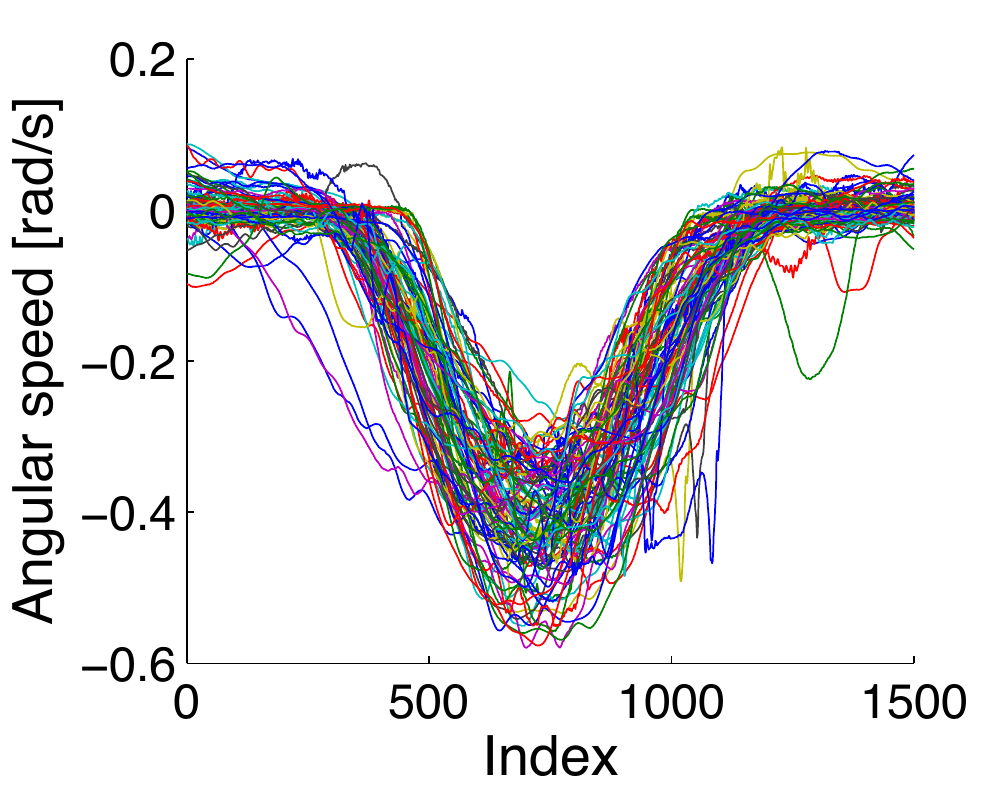}}
		\vspace{-1mm}
		\centerline{(a) Left turn.}
	\end{minipage}
	\quad\quad\quad
	\begin{minipage}[t]{0.19\textwidth}
		\centerline{\includegraphics[width=4.0cm]{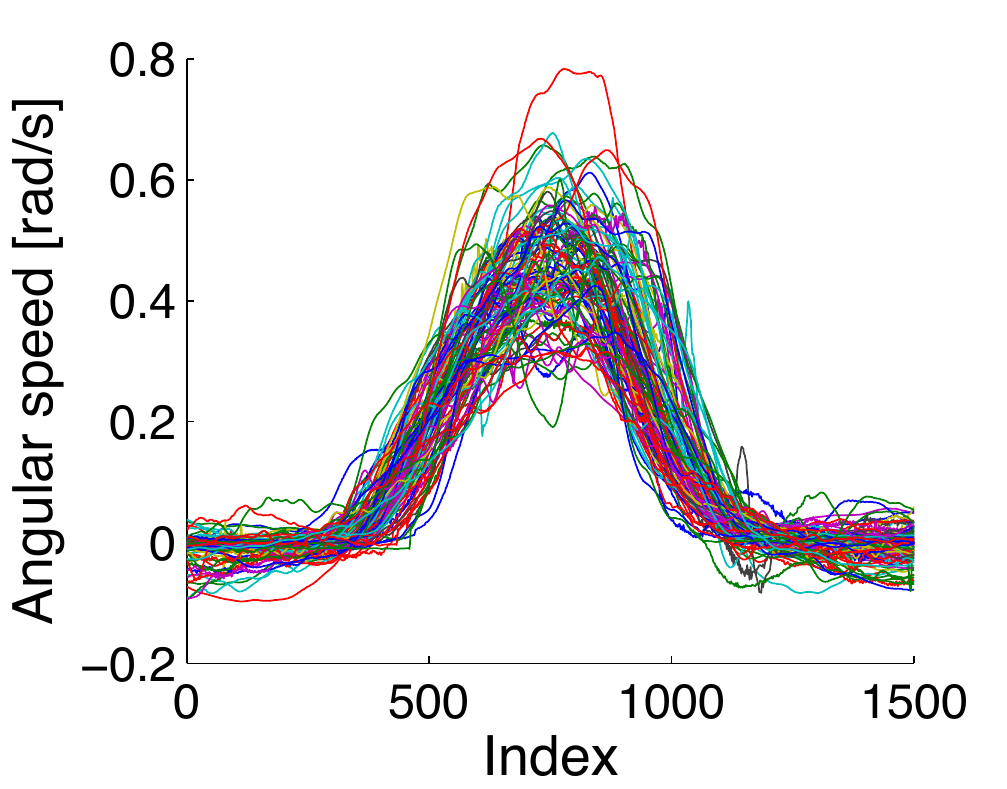}}
		\vspace{-1mm}
		\centerline{(b) Right turn.}
	\end{minipage}
	\caption{Interpolated gyroscope readings.}
	\vspace{-1em}
	\label{fig:interpolation_show}
\end{figure}

As a result, whenever the driver makes a left/right turn, \name\ can acquire a
sensor data stream (i.e., gyroscope and accelerometer readings) which was
output only during the turn, i.e., during $s = [s_{start}, s_{end}]$.
However, since different road geometries may result in different turning
radii, the length of the readings may vary, which may affect the performance of
\name. Thus, in order to make \name's fingerprinting accuracy
independent of path selection and only \textit{driver-dependent}, we
interpolate the sensor data stream into a fixed length. This also facilitates
\name\ to fingerprint the driver even when using two different devices that
may have different sampling rates.

Fig.~\ref{fig:interpolation_show} shows the gyroscope readings of
12 different drivers' left and right turns after interpolation; we will later
elaborate in Section~\ref{sec:evaluation} on how we collected them.
Near-equivalent shapes of the gyroscope readings indicate that via
interpolation, the analyses can be done from a consistent vantage point, despite
turns being made on different road geometries.
We will later show through evaluations that since the variance in left turn
radii is usually much higher than that in right turns --- as right turns
usually start from only one lane --- without such an interpolation,
\name's fingerprinting accuracy drops more when using left-turn data
than when using right-turn data.


%% file: body/system02.tex
\subsection{Maneuver-based Fingerprinting}
\label{subsec:man_fingerprint}
Whenever driver $x$ (whose identity is not yet determined) makes a left/right 
turn, \name\ acquires an IMU sensor data stream $\varphi^{x}_{turn}$.
The main challenge in fingerprinting $x$ is determining which features to extract from the data stream.

{\bf Feature extraction.}
When drivers make either a left or right turn, one might notice that some
drivers have their unique pattern in making the turn. Capturing such a pattern, \name\ extracts the following 
three new features from the filtered IMU sensor data for driver fingerprinting:
\begin{itemize}
	\item[$F_1$.] Acceleration along the end-of-turn axis ($A_{eot}$);
	\item[$F_2$.] Deviation of $F_1$ ($\Delta A_{eot}$); and
	\item[$F_3$.] Deviation of the raw yaw rate ($\Delta Y_{raw}$).
\end{itemize}

As depicted in Fig.~\ref{fig:headingaxis}, we define the start-of-turn (SOT)
axis as the axis/direction in which the vehicle was detected to have started
its turn (direction at time $s_{start}$). In reference to the SOT axis, we
define the {\em end-of-turn} (EOT) axis as the one orthogonal to the SOT axis.
That is, regardless of the change in the vehicle's heading angle after the turn (e.g., 95$^\circ$
for a right turn), by definition, the EOT axis is set perpendicular to the SOT axis.

\begin{figure}
	\centering
	\includegraphics[width=0.85\linewidth]{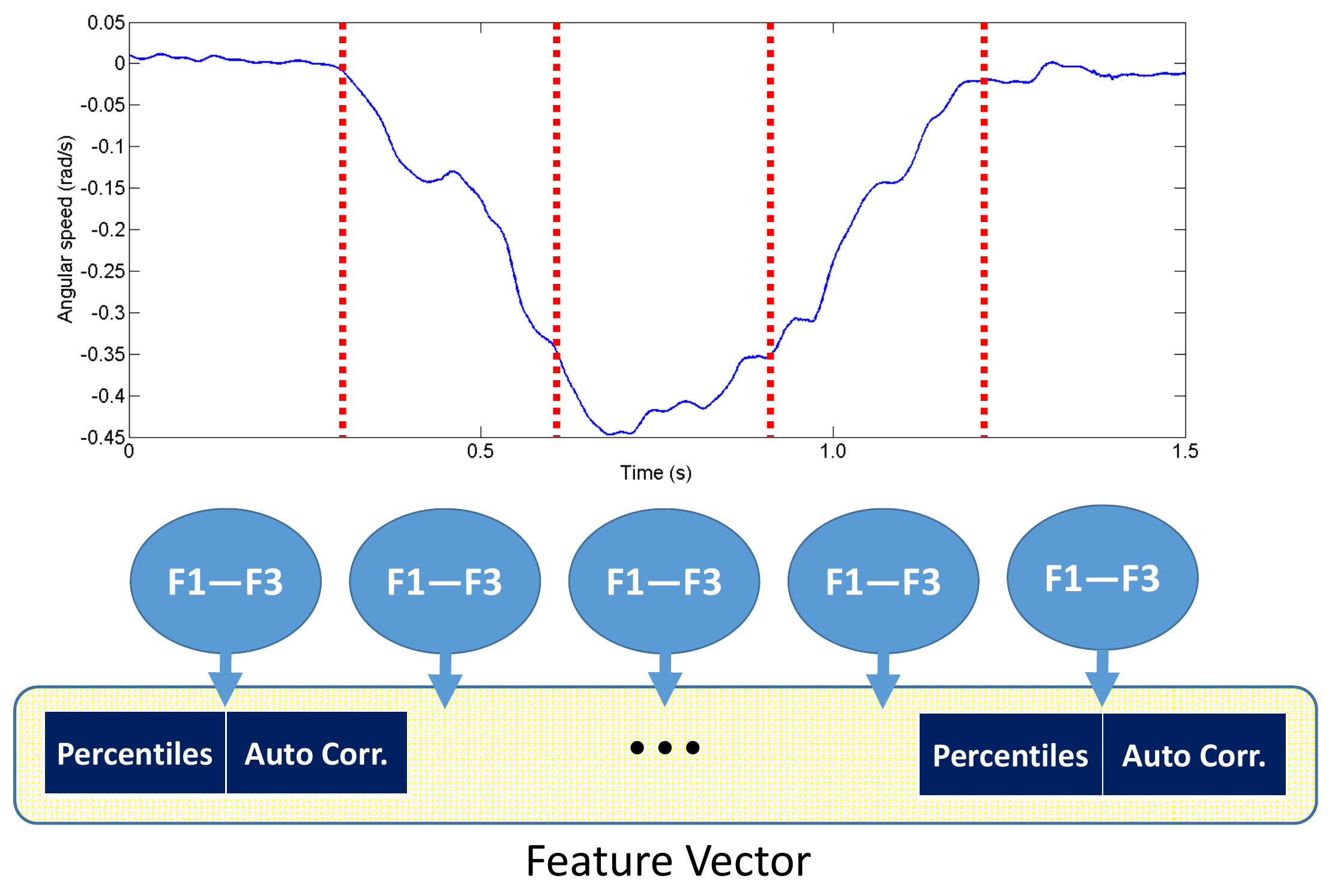}
	\vspace{-1em}
	\caption{\name's construction of feature vector.}
	\vspace{-1em}
	\label{fig:featform}
\end{figure}

{\bf F$_1$. Acceleration along the EOT axis.}
The acceleration along the EOT axis is an
interesting yet powerful feature in \name\ since it represents {\em both}
1) how much the driver turns his/her steering wheel and
2) {\em at that moment} how hard the driver presses the brake/acceleration
pedal during the left/right turn.
In other words, it reflects one's (unique) turning style. We will
later show through extensive evaluations that the features we use for \name\ do
{\em not} depend on the vehicle type or route but only on the driver's unique
maneuvering style.
Note that instantaneous acceleration, which we refer to as the acceleration
along the vehicle's heading axis, measured during a turn would only reflect
the driver's input/actions on the brake/acceleration pedal but not on the steering
wheel. Similarly, the instantaneous yaw rate, i.e., the angular velocity of the
vehicle, measured from the gyroscope sensor would only reflect the driver's
actions on the steering wheel.

For deriving the vehicle's {\em acceleration along the EOT axis} when $nT_s$
seconds has elapsed since $s_{start}$, $A_{eot}[nT_s]$, \name\ utilizes
the vehicle's instantaneous acceleration, $A[nT_s]$, at that moment (obtained
from the accelerometer) and its change in the heading angle,
$\theta[nT_s]$ (extracted from the gyroscope) as:
\begin{equation}
A_{eot}[nT_s]=A[nT_s]sin(\theta[nT_s]).
\end{equation}
In addition to the acceleration along the EOT axis, the value along the SOT
axis may also be used. However, since the information \name\ would obtain from
the accelerations along the SOT axis will be redundant when
those along the EOT axis are already available, we do not consider them as
features in \name; this also reduces the feature space.

\begin{figure}
	\centering
	\includegraphics[width=0.85\linewidth]{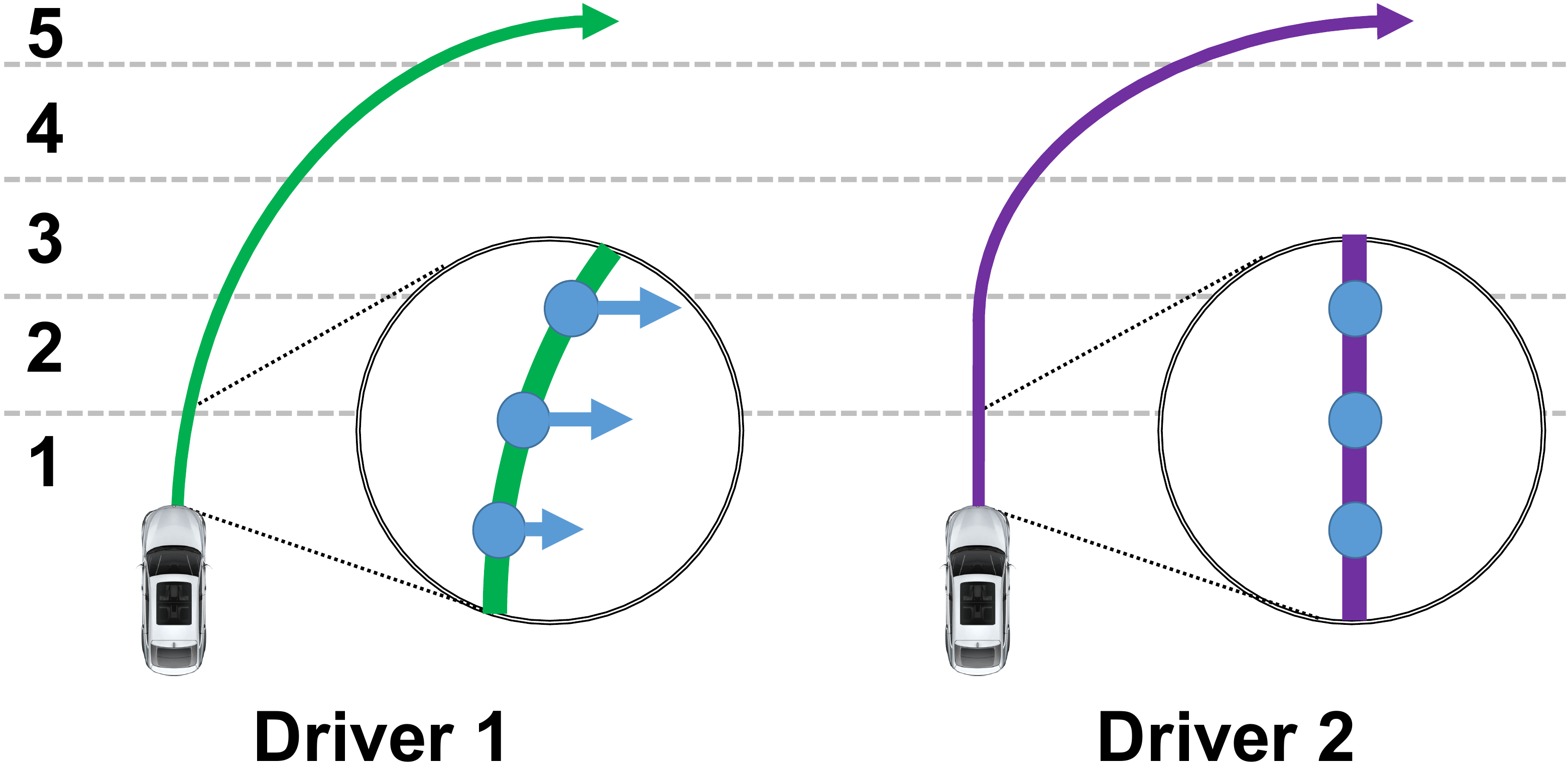}
	\caption{Different autocorrelations depending on the driver's turning
	style.}
	\vspace{-1em}
	\label{fig:autocorr}
\end{figure}

As an alternative to $A_{eot}$, one can think of using centripetal/lateral
acceleration, which would be perpendicular to the vehicle's instantaneous
acceleration ($A$). However, since the centripetal acceleration is affected by
the turning radius, whereas the acceleration along the end-of-turn axis is not,
we do not consider this for features in \name.

{\bf F$_2$--F$_3$. Deviations of $A_{eot}$ and raw yaw rate.}
\name\ derives not only $A_{eot}$ but also $\Delta A_{eot}$, i.e.,
difference between the subsequent acceleration values along the EOT axis.
Since $\Delta A_{eot}$ reflects how {\em aggressively} the driver concurrently
changes his steering and pedal actions during a turn, this feature
captures the driver's aggressiveness during the turn.

In addition to $\Delta A_{eot}$, \name\ also determines the deviations in the
{\em raw} yaw rate measurements, $\Delta Y_{raw}$. Note that in order to
accurately extract left/right turns, \name\ pre-processed the data with a
low-pass filter. However, as the turns are already extracted, in order to
not lose the accurate understanding/interpretation of how aggressively the
driver turns his steering wheel, \name\ also derives $\Delta
Y_{raw}$; the driver's aggressiveness shown from the low-pass filtered data
would have been reflected in $F_2$.
In addition to the driver's aggressiveness of turning the steering wheel, this
feature also captures how stable the driver maintains an angle during the
turn(s) and thus helps \name's driver fingerprinting.

{\bf Feature Vector Construction.}
To construct the feature vector $\gamma_{turn}$ for classification and thus fingerprinting,
\name\ transforms $F_1$--$F_3$ as follows:
\begin{itemize}
   \item[1.] Upon detection of a turn, as shown in Fig.~\ref{fig:featform},
	\name\ divides the IMU sensor measurements (acquired during the turn) into
	5 stages, each with an identical duration.
    \item[2.] For each stage, \name\ determines $F_1$--$F_3$.
    \item[3.] For each of $F_1$--$F_3$, \name\ determines its \{10, 25,
    50, 75, 90\}-th percentiles and autocorrelations at 1--10
    lags and aggregates them for constructing a feature vector.\footnote{\name\
    does not use statistics such as mean, variance, and min./max.,
    since (based on our observation) they do not help in fingerprinting the
    driver; they only increase the size of the feature space, unnecessarily.}
\end{itemize}
Note that \name\ generates an instance with such a feature vector per
(detected) turn. With the percentiles, \name\ understands the distributions
of $F_1$--$F_3$ in each stage of turn.

Meanwhile, a more interesting and powerful feature for \name\ in
fingerprinting the driver is the {\em autocorrelations} of $F_1$--$F_3$ in each
stage of turns. Fig.~\ref{fig:autocorr} shows an example of two different
drivers making a right turn.
When making the right turn, one can see that driver 1
started turning his steering wheel during stage 1 of the turn whereas driver 2
started it later during stage 3. As shown in Fig.~\ref{fig:autocorr}, which also
illustrates the accelerations along the EOT axis ($A_{eot}$) during stage 1,
one can see that an early turn from driver 1 incurs non-zero values of
$A_{eot}$ in stage 1 of the turn.
On the other hand, since driver 2 drives further on a straight line
along the SOT axis, his $A_{eot}$ values in stage 1 would approximately be 0.
Similarly, values of $F_2$ and $F_3$ would also remain 0 for
driver 2, but not for driver 1. As a result, the autocorrelations of
$F_1$--$F_3$ for driver 1 would show significantly different values from those
for driver 2, i.e., drivers' different maneuvering styles lead to
different $F_1$--$F_3$ autocorrelations during a turn.

\begin{figure}
	\centering
	\includegraphics[width=0.95\linewidth]{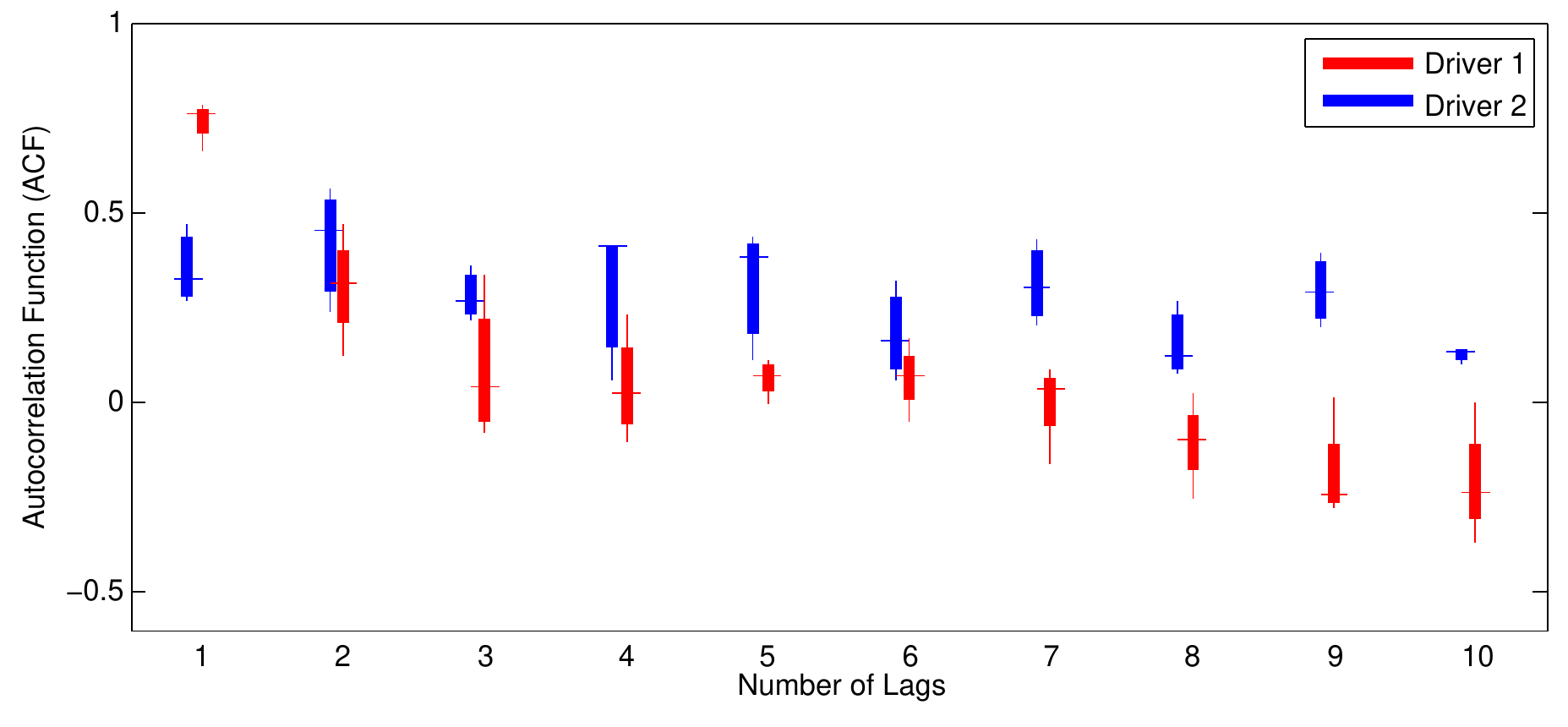}
		\vspace{-0.7em}
	\caption{Correlogram of feature $F_1$ for two drivers.}
	\vspace{-0.9em}
	\label{fig:autocorr_example}
\end{figure}

Then, are these autocorrelation values of $F_1$--$F_3$ different enough
between drivers to be considered as a driver's fingerprint?
Also, for a given driver, are those values consistent across multiple left/right turns?
Fig.~\ref{fig:autocorr_example} shows the boxplots of $F_1$ autocorrelations for two 
drivers --- who participated in our evaluations --- during their first stage of left turns.
We will later elaborate on the evaluation settings in Section~\ref{sec:evaluation}.
One can see that since the {\em tendencies} of drivers moving straight or
turning the steering wheel early/late at the early stages of turns were different, 
the autocorrelations (at different lags) between the two drivers were obviously distinguishable.
Moreover, one can see that although the driver was making those
left turns at different times and places, the variances in some autocorrelation
lags were quite low, i.e., stable.
Not only the first stage but also stages 2$\sim$5 showed a similar
distinctiveness and stability.
This shows that the autocorrelations of $F_1$--$F_3$ are not only distinct
among drivers but also quite stable for a given driver, i.e., drivers'
turning styles are relatively constant and distinct, so as to function as
the core for \name\ in fingerprinting the drivers.

{\bf Driver fingerprinting after a single turn.}
Using the constructed feature vector $\gamma^{x}_{turn}$ as an input and 
driver\textunderscore 
profile\textunderscore table as the training dataset for machine classifiers
(e.g., Random Forest, SVM, Naive Bayes), \name\ can fingerprint the 
driver as soon as the driver has made either a left or right turn, which we 
refer to as a ``maneuver-based approach''. 
As shown in Fig.~\ref{fig:overview}, driver\textunderscore 
profile\textunderscore table stores the 
drivers' identities (e.g., Alice, Bob) and their corresponding driving data, $\Gamma_{turn}$. 
For now, we assume that the targeted driver/victim is 
within driver\textunderscore profile\textunderscore table, and this table is 
given to the adversary. 
Note that existing approaches for driver fingerprinting also assume this.
Since this does not hold in some practical circumstances, we will discuss in  
Section~\ref{subsec:trainset} how the adversary may construct/obtain 
driver\textunderscore profile\textunderscore table 
from scratch via unsupervised machine learning.

%
\vspace{-0.5em}
\subsection{Trip-Based Fingerprinting}
\label{subsec:trip_fingerprint}
Albeit quite effective, when trying to fingerprint the driver within just one
turn, some false positives/negatives may occur, possibly due to a sudden change
in traffic signals, interruptions from pedestrians, etc. Hence, in order to
reduce/remove such false positives/negatives, \name\ can exploit the 
``accumulated'' data obtained from multiple left/right turns within a trip that 
the driver is making, i.e., trip-based approach.
Note that during a trip the driver remains the same.

One way the adversary might achieve this via \name\ is by exploiting the Naive
Bayes classifier, which is a simple probabilistic classifier based on the
Bayes' theorem. For a given vehicle driven by $N$ different drivers,
assume that the adversary has a training set composed of several instances
labeled as one of ${{D_1},\cdots,{D_N}}$.
Then, within the trip in which the adversary attempts to fingerprint the
driver, as the driver makes more turns, i.e., as more instances are
collected, the adversary can estimate the maximum posterior probability
(MAP) and thus predict the driver to be $D_{pred}$ as:
\begin{equation}\label{eqn:map}
D_{pred}=\mathop {\arg \max }\limits_{k \in \left\{ {{\rm{1,}} \cdots ,N}
\right\}} p\left( {{D_k}} \right)\prod\limits_{i = 1}^n {p\left( {{T_i}|{D_k}}
\right)},
\end{equation}
\noindent
where $n$ is the number of vehicle turns made up to the
point of examination during the trip. Here, $p(T_i | D_k)$ represents the
likelihood that the (measured) $i$-th turn, $T_i$, would have occurred, given
driver $D_k$ is driving the vehicle.
Even though the adversary assumes that the prior probability, $p(D_k)$ is
equivalent across the (potential) drivers, i.e., each driver has an equal
probability of driving that vehicle, we will later show through evaluations that the
adversary can fingerprint the driver with higher accuracy than just using one
turn, although, in most cases, one turn was sufficient in correctly
fingerprinting the driver.

\vspace{-0.5em}
\subsection{Training Set Formulation}\label{subsec:trainset}
\begin{figure}
	\centering
	\includegraphics[width=\linewidth]{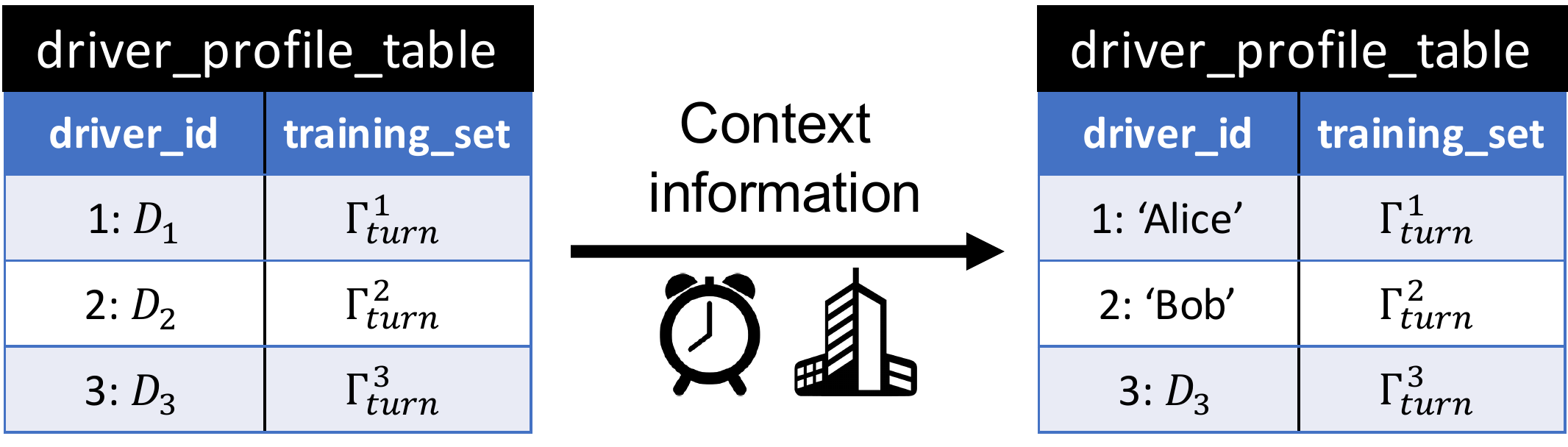}
	\vspace{-1em}
	\caption{Formulating the training set (driver\textunderscore 
		profile\textunderscore table) from scratch.}
	\vspace{-1em}
	\label{fig:form_training_set_01}
\end{figure}
We now present how to construct driver\textunderscore profile\textunderscore table from 
scratch, thus enhancing {\name}'s practicability. 
In constructing a valid training set, there exist two main challenges: 
1) determine whether the new collected data comes from a driver the 
adversary had already learned about and 
2) for the labeled data, exactly know who that driver would be.
For the latter, without knowing this, the adversary would only be able to label 
the data as some variable $D_i$ rather than the driver's actual identity 
(e.g., $D_1$=``Alice''); see Fig.~\ref{fig:form_training_set_01} (left).

{\bf Determining a new driver.}
As illustrated in Fig.~\ref{fig:overview}, once the adversary equipped with \name\
collects $\gamma^{x}_{turn}$ from driver $x$, {\name} determines whether driver 
$x$ would be one of the known/learned drivers or a new (unknown/unlearned)
driver, i.e,. whether $x$ belongs to driver\textunderscore 
profile\textunderscore table. 
In the former, the adversary can expand the existing
training set, whereas in the latter, he would have to construct a new training
set for driver $x$. Such a process is essential, especially when the
adversary starts, for the first time, to fingerprint the driver of a vehicle, i.e.,
starting from scratch.

Here, we briefly discuss how the adversary can indeed utilize
unsupervised machine learning to correctly cluster/label $\gamma^{x}_{turn}$ to
either an already-known or a new driver. What the adversary may do is label
$\gamma^{x}_{turn}$ based on its log-likelihood obtained from a Gaussian
mixture model (GMM). GMM is a combination of Gaussian component densities that
are used for modeling the probability distribution of continuous measurements;
see \cite{gmm} for more details.

Suppose the adversary starts to fingerprint the driver(s).
At first, since he has an empty training set, he first builds a GMM
model, $M_1$, based on the training data $\Gamma^{1}_{turn}$ acquired during the vehicle's first trip
and labels it as (some) driver $D_1$. Then, during the next trip, when the
adversary acquires $\gamma^{x}_{turn}$, he calculates the log-likelihood of
$\gamma^{x}_{turn}$ given $M_1$.
Accordingly, if the log-likelihood is high, meaning that
$\gamma^{x}_{turn}$ is likely to be output from driver $D_1$, \name\ appends
$\gamma^{x}_{turn}$ to the associated training set $\Gamma^{1}_{turn}$. On the other hand, if
the log-likelihood is low, $\gamma^{x}_{turn}$ is likely to have been generated
by a new driver $D_2$ . Accordingly, \name\ makes a new training set 
$\Gamma^{2}_{turn}$. 
In such a way, the adversary would able to construct the initial version of 
driver\textunderscore profile\textunderscore table (e.g., 
Fig.~\ref{fig:form_training_set_01} (left)).

{\bf Accurately labeling the data.}
As shown in Fig.~\ref{fig:form_training_set_01}, the adversary can construct the training dataset more concretely
if he knows exactly who $D_1$ is (e.g., $D_1$=``Alice''). This can be achieved
not only via oversight but also based on other context information.
For example, if the adversary knows that Alice always drives to work for an 
hour at 9:00 am on weekdays, the data being collected during 9:00$\sim$10:00 am 
is more likely to reflect Alice's rather than other's driving behavior.
Other than time, useful context information such as location, phone usage 
patterns, DNS traffic pattern may also be utilized in verifying that the driver 
is indeed ``Alice''. Note, however, that these are only valid when that context 
becomes available and is valid (e.g., on weekends, we are not sure that Alice 
is driving). Thus, we use context only for constructing the training dataset of 
\name, but not for the actual fingerprinting.

In fact, such an approach would not only make the adversary build a concrete
training set but also let him estimate the prior probability of a
driver driving the vehicle --- $p(D_k)$ in Eq.~(\ref{eqn:map}) --- and thus
increase the fingerprinting accuracy.

In Section~\ref{sec:evaluation}, we will show that the adversary can 
construct/obtain a well-formulated training set via this GMM approach.
We also show through extensive evaluations that even when the training dataset
obtains a few instances with incorrect labels (\ref{eval:erroneous_data}), i.e., a (slightly) defective
training set due to the adversary's mistake, he may still be able
to identify the driver with high accuracy.

%


%
%
%

%% file: body/evaluation.tex
\section{Evaluation}
\label{sec:evaluation}
\input{body/eval01}
\input{body/eval02}

%% file: body/eval01.tex
\subsection{Evaluation Setup}
\label{subsec:eval_setup}
To validate whether \name\ accurately fingerprints the {\em driver}, we 
evaluated two imperative aspects of \name: whether the derived features 
of \name\ are
1) dependent only on the driver and
2) remain constant with various changing factors, e.g. car and/or route.
To validate these, we started from a small-scale experiment where we varied/controlled 
different factors such as driver, car, route, which may (or may not) affect 
\name's performance.
Once validated, to evaluate how \name's performance scales with high number of 
drivers, we conducted a large-scale experiment with more drivers driving in 
their own choice of car and routes.
Overall, our driving data collection took 4 months and had more than 116 hours 
of driving data obtained from urban/suburban areas. The accumulated driving 
milage was 1,688 miles.


\begin{figure}[t]
\centering
	\begin{minipage}[t]{0.24\textwidth} 
		\centerline{\includegraphics[width=4.36cm]{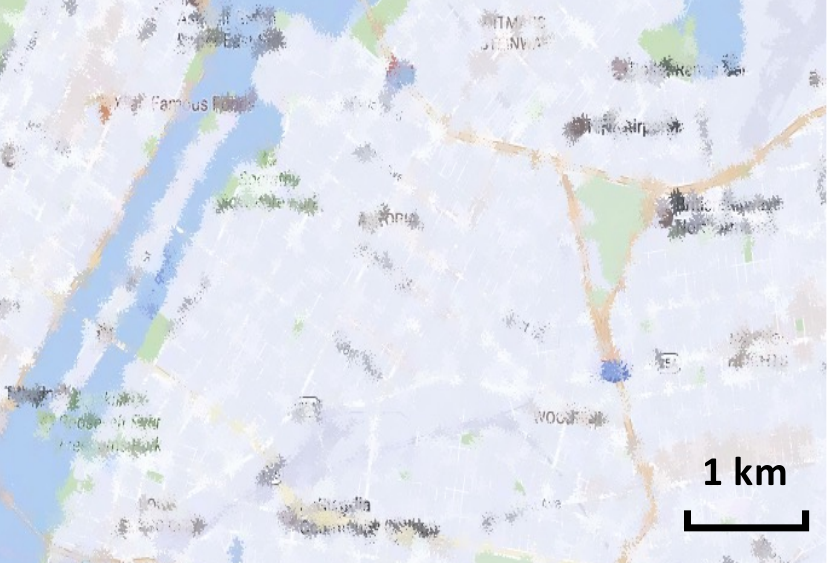}}
		\centerline{(a) Metropolitan area.}
	\end{minipage} 
	\quad
	\begin{minipage}[t]{0.19\textwidth}
		\centerline{\includegraphics[width=3.36cm]{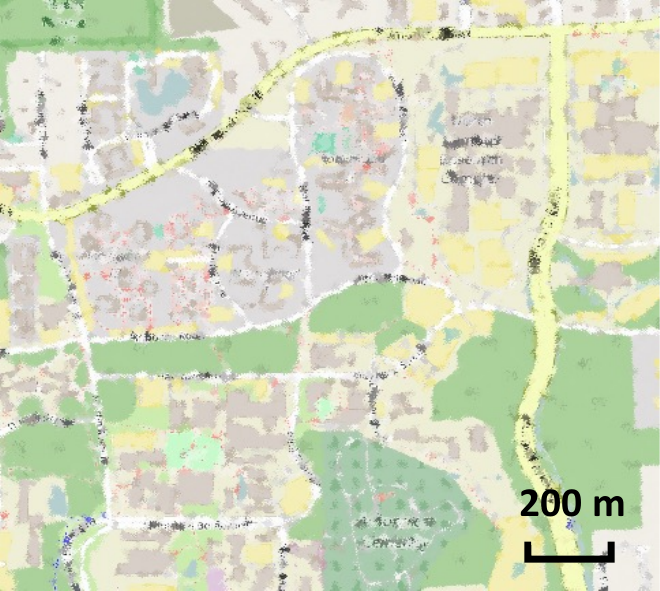}}
		\centerline{(b) Suburban area.} 
	\end{minipage} 
	\caption{Different driving environments.}
	\label{fig:route_select}
	\vspace{-1em}
\end{figure}
\textbf{Data-collection methodology.}
The data-collection module of {\name} was implemented as an Android app and was installed on various 
Android smartphones including Google Pixel, Nexus 5X, Samsung Galaxy S5, and Samsung Note 5. 
We recruited 12 (9 male and 3 female) drivers with an age span of 22--50. 
For collecting the data, we asked the drivers to 1) ensure the app is turned on 
before driving and 2) upload the data after finishing a trip. 
According to our survey after data collection, none of the participants 
indicated that our data collection app affected their normal driving pattern.

Since our system does not require any personal information from the users, the Institutional 
Review Boards (IRB) of our university classified this effort as non-regulated. 

\textbf{Tested vehicle and routes.}
The selection of vehicle and road were only controlled in our small-scale 
experiment.
Specifically, to verify the factors which affect \name's 
performance, we asked two recruited drivers to drive a Honda Sedan and a Ford 
SUV. 
In the large-scale experiment, to validate that the derived features in \name\ 
and thus its fingerprinting  do not depend on the vehicle of choice, we allowed 
all participants to drive their own vehicle(s). 
As a result, we collected data from 10 cars of 7 different models: 
Honda Accord Sedan, 
Honda CRV SUV, Toyota Camry Sedan, Ford Explorer SUV, Hyundai Elantra Sedan, Jeep Compass SUV, and Toyota Corolla Sedan.
Moreover, the routes were also freely chosen by the driver which included those 
in a suburban area with less traffic or a metropolitan area with heavy traffic; 
examples shown in Fig.~\ref{fig:route_select}.

%

%% file: body/eval02.tex
\vspace{-0.5em}
\subsection{Factor Analysis \& Experimental Design}
\label{subsec:eval_factors}
To verify that the fingerprinting accuracy of \name\ only depends on the driver, not 
on the car or route, we first conducted a factor analysis via a small-scale 
experiment. As shown in Table~\ref{tab:eval_summary}, we conducted 6 
experiments, T1--T6, with same/different drivers, cars, and/or routes.
For T7 (the large-scale experiment) where every factor was 
varied and uncontrolled, we will discuss later on how \name\ performed.


\begin{table}
\centering
	\begin{tabular}{llc}
		\toprule
		{\bf Differentiated Factor(s)} & {\bf Constant Factor(s)} 
		& \bf{Acc.}\\
		\midrule
		T1. Car & Driver, Route &  Low\\ 
		T2. Route & Driver, Car & Low\\
		T3. Car, Route & Driver & Low \\
		T4. Driver& Car, Route & High \\
		T5. Driver, Car & Route & High \\ 
		T6. Driver, Route & Car & High \\ 
		T7. Driver, Car, Route & (None) & High \\
		\bottomrule
	\end{tabular}
	\vspace{5pt}
	\caption{Summary of evaluations.}
	\label{tab:eval_summary}
	\vspace{-3em}
\end{table}

{\bf Test cases.}
For tests T1--T6, we varied/controlled the three factors as follows:
\begin{itemize}
	\item {\it Driver Factor}: For test cases T4--T6 where the driver was 
	differentiated, two different drivers were asked to drive a 
	same/different car with specified instructions when needed, e.g., whether 
	to drive on a pre-determined route.
	\item {\it Car Factor}: For test cases T1, T3, and T5 in which the car type
	was varied, we used two different cars: Honda Accord Sedan and a Ford 
	Explorer SUV. 
	\item {\it Route Factor}: For test cases T1, T4, and T5, where the route 
	was fixed, we asked the drivers to drive around campus along the 
	pre-determined route. For other test cases (T2, T3, and T6) where the route 
	was differentiated, the route was solely determined by the drivers.	
\end{itemize}

If \name's constructed features only depend on the driver factor, i.e., 
dependent on only the driver's unique turning style, \name's performance 
in test cases T1--T3 would be low whereas in T4--T6, it should be high.

{\bf Classification.}
For each test case, we acquired data via \name\ from two different trips, 
which differ in driver/car/route or a combination thereof (as shown in Table~\ref{tab:eval_summary}).
As the two trips (per test case) have distinct factors, we 
labeled the vehicle turns based on which {\em trip} they occurred. 
For example, in T1 where ``car'' was the only differentiated factor between 
the two trips, although the driver was identical, the vehicle turn data from 
each trip were labeled differently as 0 and 1, i.e., binary. 
Similarly in T6 where the ``driver'' and ``route'' were the differentiated 
factors, turns from each trip were again labeled 0 and 1.
Based on the collected data from the two trips of cases T1--T6, we 
trained the classifiers using 90\% of the turns 
and leave the remaining 10\% as the test set. To obtain an accurate estimate of 
the model prediction performance, we used 10-fold cross validation.
For each test case, as turns were from two different trips (with different 
drivers/cars/routes), we used binary classification.
The classifiers we used for testing T1--T6 were Support Vector Machine (SVM) 
and a 100-tree Random Forest.

\begin{figure}
	\centering 
	\includegraphics[width=0.95\linewidth]{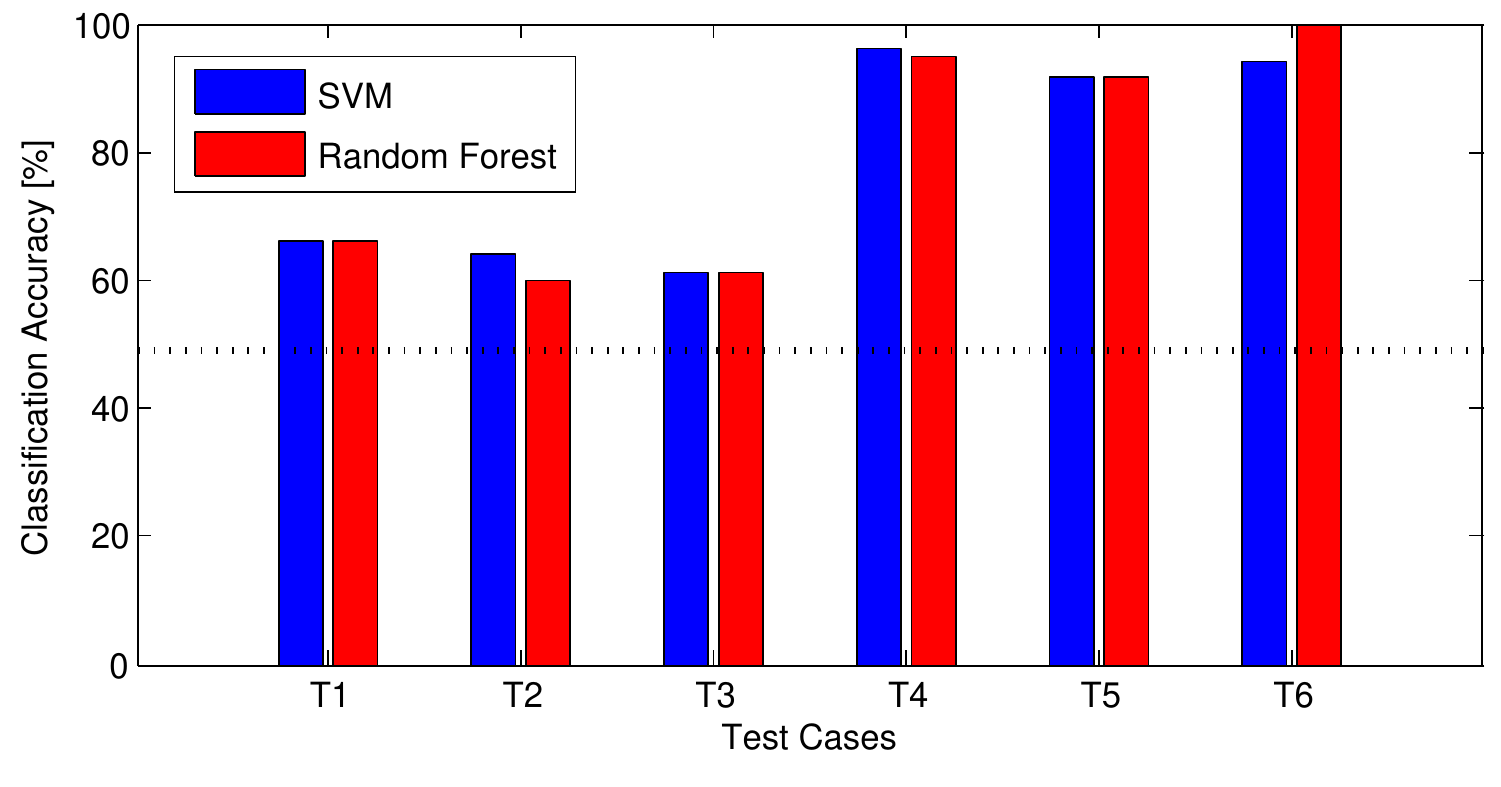}
	\vspace{-1em}
	\caption{Classification accuracy for 
	test cases T1--T6 using SVM and Random Forest.}
	\label{fig:maneuver}
	\vspace{-1em}
\end{figure}

{\bf Performance.}
Fig.~\ref{fig:maneuver} plots accuracies of \name\ in fingerprinting the driver
based one single turn in T1--T6, when using SVM and Random Forest.
Since the classification only needed to be binary, the classification accuracy of random guessing is 50\%, which is shown as a horizontal dotted line.

One can see that for test cases T1--T3, although the vehicle and/or the route were 
different, \name\ showed a very low classification accuracy:
66.6\%, 64.2\%, 61.1\% using SVM, and 66.6\%, 60.4\%, 61.1\% using Random 
Forest in cases T1--T3, respectively. 
Such a result can, in fact, be interpreted as having a similar accuracy as when 
it is guessed randomly. 
This also implies that regardless of the car or route used/taken, if the driver 
is identical, \name\ gets {\em confused}.

When the "driver" factor was changed as in test cases T4--T6, one can see from 
Fig.~\ref{fig:maneuver} that the classification accuracy of \name\ was much 
higher: 96.3\%, 91.7\%, 94\% using SVM, and 95\%, 91.7\%, 100\% using Random 
Forest in cases T4--T6, respectively. 
Such a high classification accuracy was due to the fact that 
between the two trips of T4--T6, the drivers were different.

Based on these results, we can conclude that the features derived by \name\ 
 depends {\em only} on the driver and not on 
other factors such as car and/or route, thus functioning as the key for accurate driver fingerprinting. 
Moreover, {\name} shows consistent performance across different  
machine classifiers.

\vspace{-0.5em}
\subsection{Large-scale Experiment}
To further evaluate {\name}'s performance with more 
drivers, and to verify whether its derived features for a given 
driver remain consistent across different routes, we conducted a large-scale 
experiment: we used all the sensor data acquired from the 12 participants who 
drove 10 different cars. As most of these participants drove different cars on 
different routes, test case T7 represents such a setting.

In T7, since there were more than 2 drivers, when using SVM and Random Forest, 
we performed a multi-class classification. To achieve this, we examined it 
through one~vs.~one reduction rather than one~vs.~all since the former reflects 
more accurate results than the latter~\cite{onevsone}. In the dataset, feature 
vectors of turns were labeled depending on who the driver was.
Again, 10-fold cross validation was performed for an accurate performance 
measure.

\begin{figure}
	\centering 
	\includegraphics[width=0.95\linewidth]{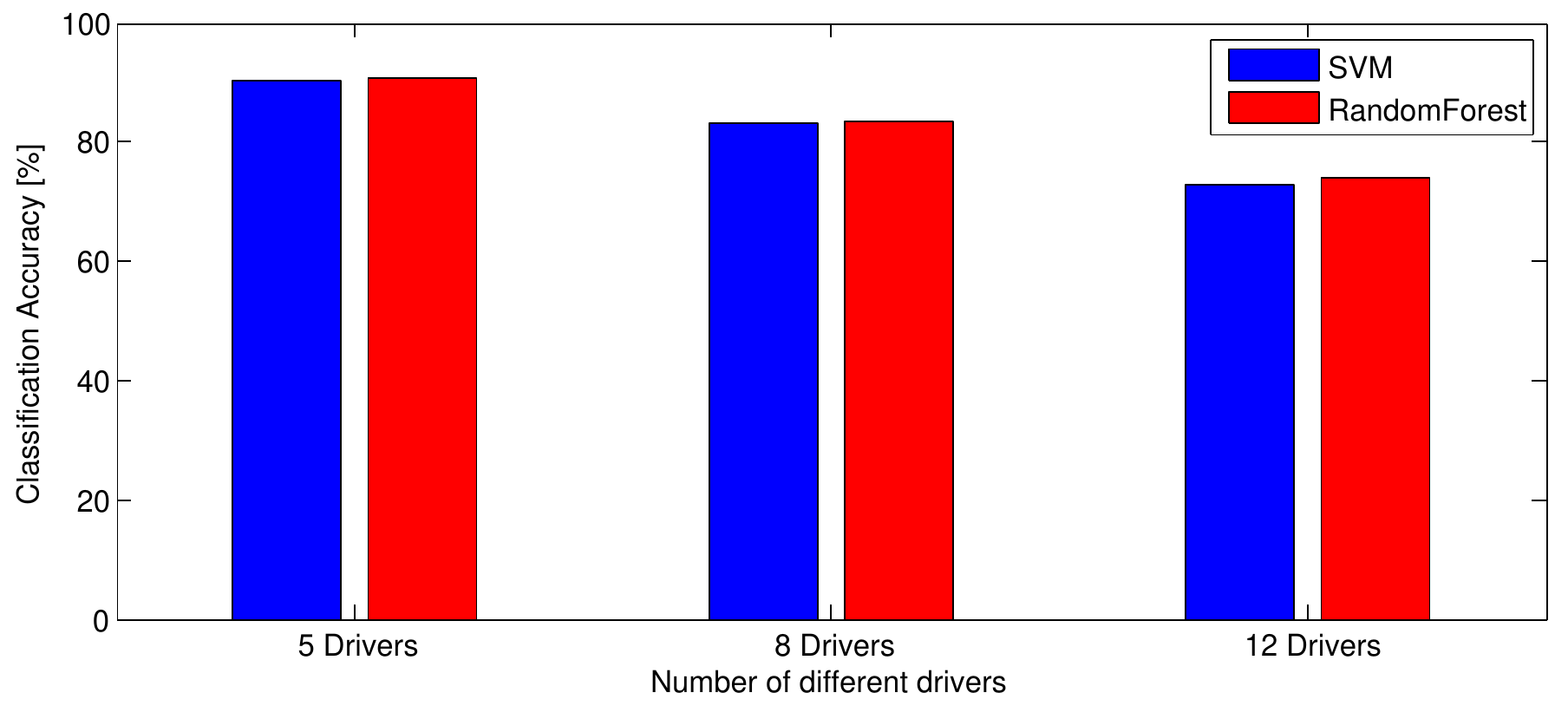}
	\vspace{-1em}
	\caption{\name's accuracy in fingerprinting 5, 8, and 12 drivers within one 
	vehicle turn using SVM and Random Forest.}
	\vspace{-1em}
	\label{fig:largescale}
\end{figure}

\textbf{Maneuver-based approach.}
We first evaluate how well \name\ identifies 5, 8, and 12 drivers using a 
maneuver-based approach. 
Fig.~\ref{fig:largescale} plots \name's accuracy in fingerprinting 5, 8, 
and 12 different drivers using SVM and Random Forest. 
One can see that within only one left/right turn, \name\ can fingerprint the 
driver with 90.5\%, 83.1\%, and 72.8\% accuracies across 5, 8, and 12 
drivers, respectively, using SVM. When Random Forest is used, the fingerprinting 
accuracies were shown to be 90.8\%, 83.5\%, and 74.1\% across the same driver sets. Although only mobile IMU sensors were used by \name, 
thanks to its new features, \name\ was able to identify the driver even 
though the number of drivers got larger; much better than random guessing. 
Such an achievement was made by observing only {\em one} left/right turn! 

\textbf{Trip-based approach.}
As discussed in Section~\ref{subsec:trip_fingerprint}, instead of trying to 
fingerprint the driver based on one turn, the adversary may attempt to do it 
by accumulating sensor data of multiple turns collected within the trip, i.e., 
trip-based approach. To evaluate how well an adversary exploiting \name\ may 
fingerprint the driver with such an approach, we evaluated \name\ as follows.
Per iteration, from our 12-driver driving dataset, we randomly selected one 
{\em trip} made by some driver; each driver made at least 2 trips. Then, we 
first randomly permuted the vehicle turns made within that trip and then 
considered those as our test set. 
Vehicle turns made in all other trips were considered as our training set. 
In predicting who the driver was in the (randomly) selected trip, 
i.e., the driver of the test set, we used the Naive Bayes classifier, which 
predicts the label based on the maximum {\em a posteriori} (as in 
Eq.(\ref{eqn:map})). The prior probability was set to be uniform.
We performed such an evaluation 500 times.

\begin{figure}
	\centering 
	\includegraphics[width=0.9\linewidth]{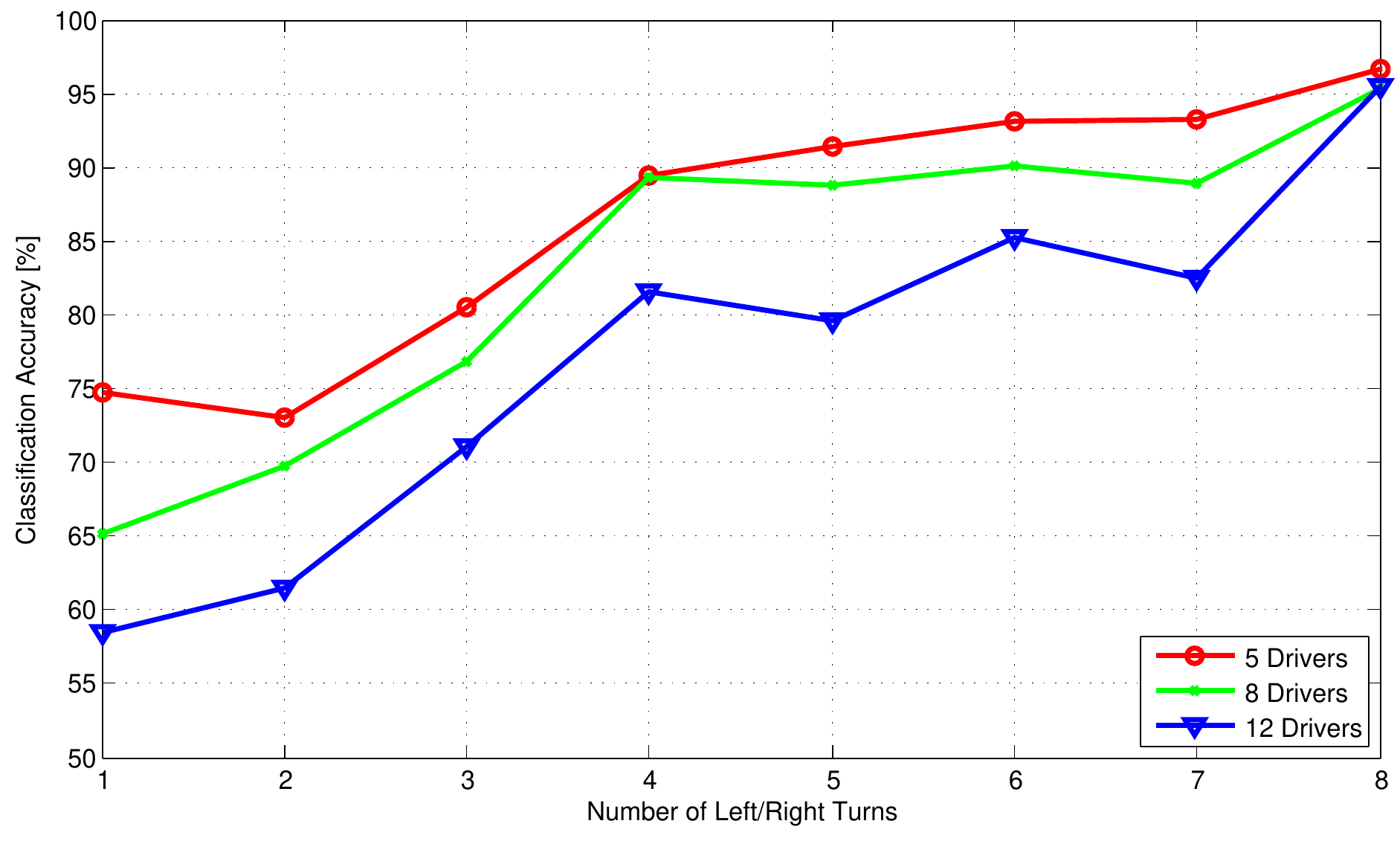}
	\vspace{-1em}
	\caption{Classification accuracy of trip-based approach using Naive Bayes.}
	\label{fig:tripbased}
	\vspace{-1em}
\end{figure}

Fig.~\ref{fig:tripbased} plots \name's accuracy in identifying the driver 
correctly for the 500 iterations using a trip-based approach, when the number of
candidate drivers were 5, 8, and 12. For evaluating the first two cases with 5 
and 8 drivers, per iteration, they (as well as their trip/turn data) were 
randomly chosen from the total of 12 drivers.
One can see that as more left/right turns were observed and analyzed 
by \name, its classification accuracy continuously increased.
After observing 8 left/right turns, \name\ achieved fingerprinting accuracies 
of 96.6\%, 95.4\%, and 95.3\% across 5, 8, and 12 drivers, respectively, which 
obviously is a great improvement over the ``maneuver-based approach'', 
i.e., fingerprinting after one left/right turn.
Since the way the drivers made their left/right turns was 
{\em occasionally} inconsistent, one more turn made by the driver did not 
necessarily increase \name's performance, i.e., performance did not 
monotonically increase. However, since the drivers made most of their turns
according to their {\em usual} tendency/habit, ultimately the accuracy improved. 
Note that the accuracy of fingerprinting the driver via Naive Bayes after only 
one turn was a bit lower than when using other classifiers such as SVM or 
Random Forest due to its (naive) independence assumptions.

\subsection{Efficacy of Interpolation}
As discussed in Section~\ref{subsec:turnextract}, to make \name's 
fingerprinting as independent as possible from the road geometry in which the 
turns are made, we interpolate the data to a fixed length. To evaluate the 
efficacy of such an interpolation, we evaluated \name's accuracy across 
12 drivers when not executing such an interpolation. 

\begin{table}
\centering\setlength\tabcolsep{.26em}
	\begin{tabular}{lcccc}
		\toprule
		& \multicolumn{2}{c}{\bf Left Turn} & \multicolumn{2}{c}{\bf 
			Right Turn} \\
		& {\bf SVM(\%)} & {\bf RF(\%)} & {\bf SVM(\%)} & {\bf RF(\%)} \\
		\midrule
		w/ Interpolation & 73.1 & 78.0 & 74.1 & 74.3\\
		w/o Interpolation & 65.2 & 72.0 & 71.5 & 72.2\\
		Average difference & \multicolumn{2}{c}{-6.95} & 
		\multicolumn{2}{c}{-2.35}\\
		\bottomrule
	\end{tabular}
	\vspace{5pt}
	\caption{Efficacy of data interpolation.}
	\label{tab:interpolation}
	\vspace{-1em}
\end{table}

\begin{figure}[t]
\centering
	\begin{minipage}[t]{0.22\textwidth} 
		\centerline{\includegraphics[width=3.60cm]{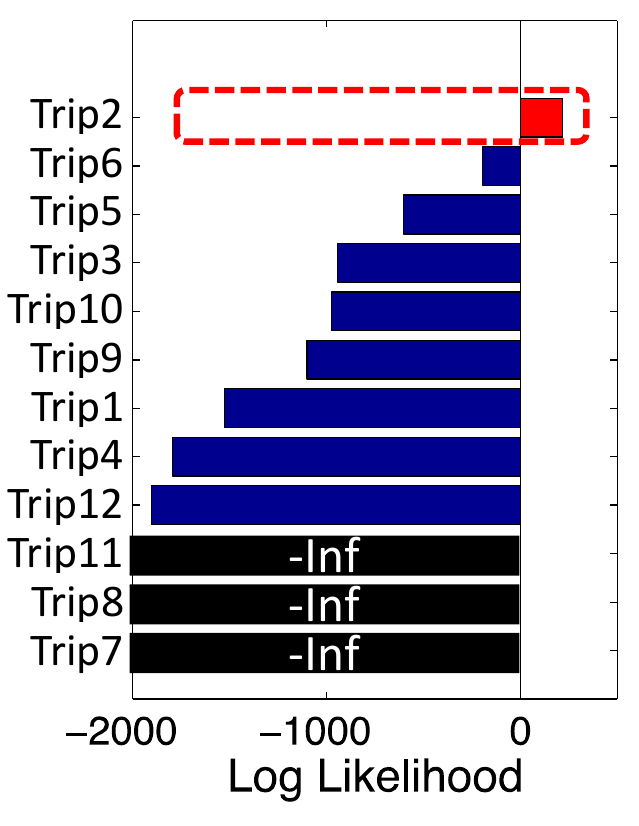}}
		\centerline{(a) Learned driver.}
	\end{minipage} 
	\begin{minipage}[t]{0.22\textwidth}
		\centerline{\includegraphics[width=3.66cm]{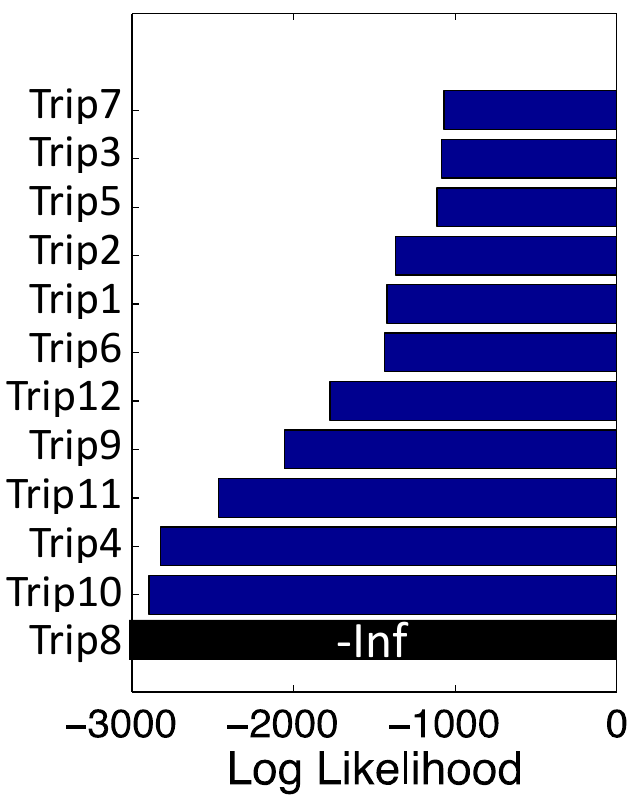}}
		\centerline{(b) New driver.} 
	\end{minipage}
	\caption{Using GMM for training set formulation.}
	\vspace{-1em}
	\label{fig:training_set_form}
\end{figure}

Table~\ref{tab:interpolation} summarizes how \name\ performed when 
fingerprinting the 12 drivers based on only left and right turns 
with/without interpolation. One can observe that when the data from different 
trips were not interpolated, the performance of \name\ dropped. The reason for 
such a drop was that road geometries for different turns (even for the same 
driver) were not identical, i.e., the turning radii 
are different. So, through interpolation, \name\ was able to remove
the possible influence of the differences in turning radii, and thus achieve 
more accurate driver fingerprinting. 
Note that a driver's turning radii can vary depending on where s/he is driving.
Here, an interesting observation is that \name's accuracy 
dropped more when identifying the driver via left turn(s) than via 
right turn(s). This was because the turning radii for left turns normally have 
higher variations between them than for right turns; left 
turns can start from multiple lanes, whereas right turns (mostly) start from 
the rightmost lane.

\subsection{Training Set Formulation}
In Section~\ref{subsec:trainset}, we discussed how the adversary may use GMM 
to construct/obtain the training dataset for driver fingerprinting from scratch. 
To validate this, we considered the following case.
Suppose that driver $D_1$ (w/o known identity) was the first to 
drive the vehicle since the adversary started to fingerprint its driver. 
Thus, the adversary constructs his initial training dataset, $\Gamma^{1}_{turn}$ 
with label $D_1$. In such a case, we examined what the GMM 
log-likelihood would be for the data collected from a new trip given $\Gamma^{1}_{turn}$.

Fig.~\ref{fig:training_set_form}(a) plots what the log-likelihood values were 
when data from 12 different trips, {\it Trip1}--{\it Trip12} (each chosen from the 12 
different drivers' trips) were considered as the test set, thus being examined 
against the GMM of $\Gamma^{1}_{turn}$. 
We constructed $\Gamma^{1}_{turn}$ based on one of driver $D_1$'s trips, 
which was not included in the 12-trip test set.
One can see that for only the data in {\it Trip2}, the log-likelihood was 
positive whereas for all others the values were negative or even negative 
infinite. This was because the driver of {\it Trip2} was $D_1$. 
Such a result shows that by observing the GMM likelihood, the adversary can 
determine whether or not the newly collected data has been output by an 
existing driver in his training dataset. In this case, in \name, the 
adversary would append the newly collected data from {\it Trip2} to its initial 
dataset, $\Gamma^{1}_{turn}$.

This time, we randomly chose another trip from our 12-driver dataset and 
considered that as the adversary's new initial training set, i.e., different 
$D_1$ and $\Gamma^{1}_{turn}$ (than the previous ones).
We again considered the test set to be composed of 12 different trip data, 
but this time, made by drivers {\em except} for the chosen $D_1$.
Fig.~\ref{fig:training_set_form}(b) plots the GMM log-likelihood values of data 
in the test set given the new $\Gamma^{1}_{turn}$. One can see that, since there were 
no trips within the test set taken by the same person as 
$D_1$, all showed negative/negative-infinite likelihoods. 
In such a case, \name\ would determine that the newly collected data was output 
by a new driver, which he had not learned about, and thus construct a new 
training dataset for that driver.

\subsection{Erroneous Training Dataset}
\label{eval:erroneous_data}
When forming the training set via GMM, the standard for clustering new data was 
whether the GMM log-likelihood is positive or not. 
However, such a threshold setting may not always be reliable. 
Thus, to understand and evaluate how \name's performance will be 
affected when the adversary wrongly labels a turn while constructing the 
training dataset, e.g., a turn was made by driver 1 but the adversary labels it 
as by driver 2, from our dataset of 5 drivers, we arbitrarily picked and labeled 
some turns to be made by {\em any} of the 5 drivers. 
The number of arbitrarily picked turns with erroneous labels were 
varied via parameter $p_{err}$, which denotes the percentage of such erroneous 
labels. For this evaluation, we present the results obtained via SVM.
\begin{figure}
	\centering 
	\includegraphics[width=0.95\linewidth]{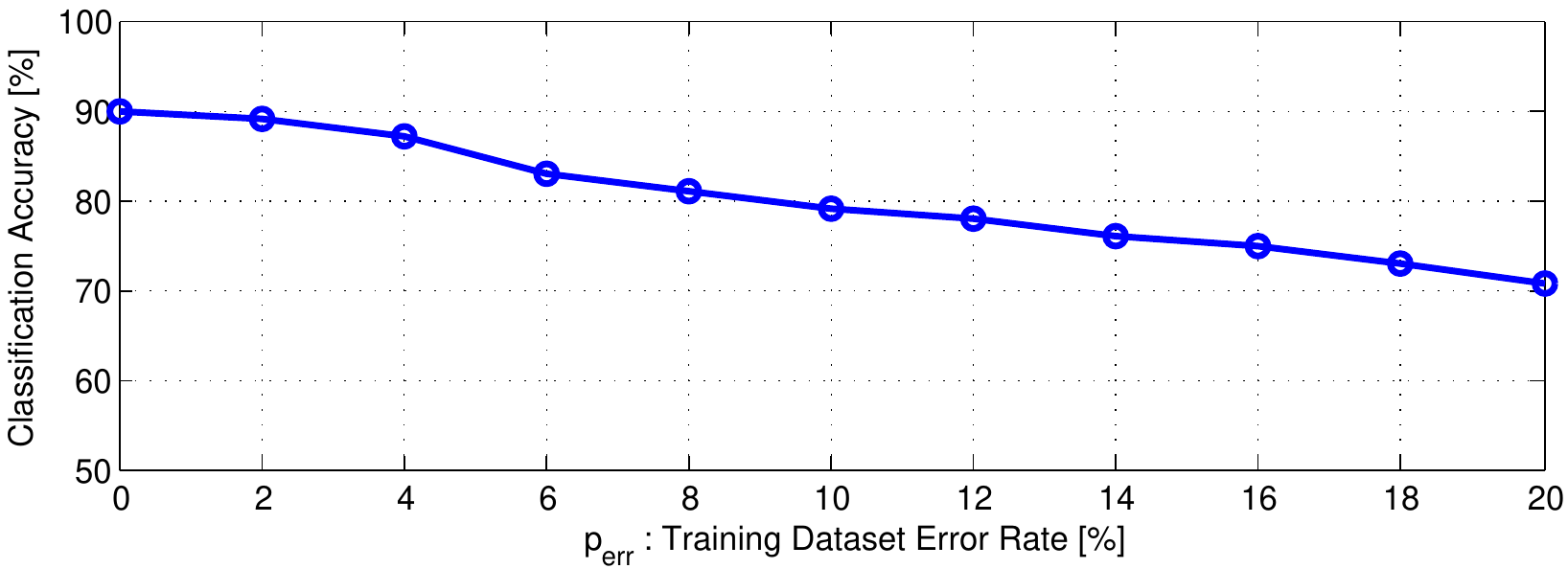}
	\vspace{-1em}
	\caption{Classification accuracy with $p_{err}$ \% erroneous training 
	dataset.}
	\vspace{-1em}
	\label{fig:errorrate}
\end{figure}

Fig.~\ref{fig:errorrate} shows how \name's fingerprinting accuracy 
changed for $p_{err}$=0$\sim$20\%. Even when the training dataset for \name\ 
contains 20\% of erroneous labels due to the adversary's mistake, 
the adversary can still achieve 70.7\% fingerprinting accuracy within 
only one turn. Despite the erroneous labels, such an accuracy can 
be increased further using a trip-based approach.
Such a result implies that the adversary may not always have to be 100\% 
accurate in constructing the training dataset in order to accurately 
fingerprint the driver with \name, which is a serious threat.

\subsection{Overhead of {\name}}
\label{subsec:overhead}
The additional overheads such as the {\em CPU usage} and {\em energy 
consumption} of \name\ on the victim's device may render the driver fingerprinting process noticeable by the victim. 

\begin{figure}[t]
\centering
	\begin{minipage}[t]{0.50\textwidth} 
		\centerline{\includegraphics[width=0.95\linewidth]{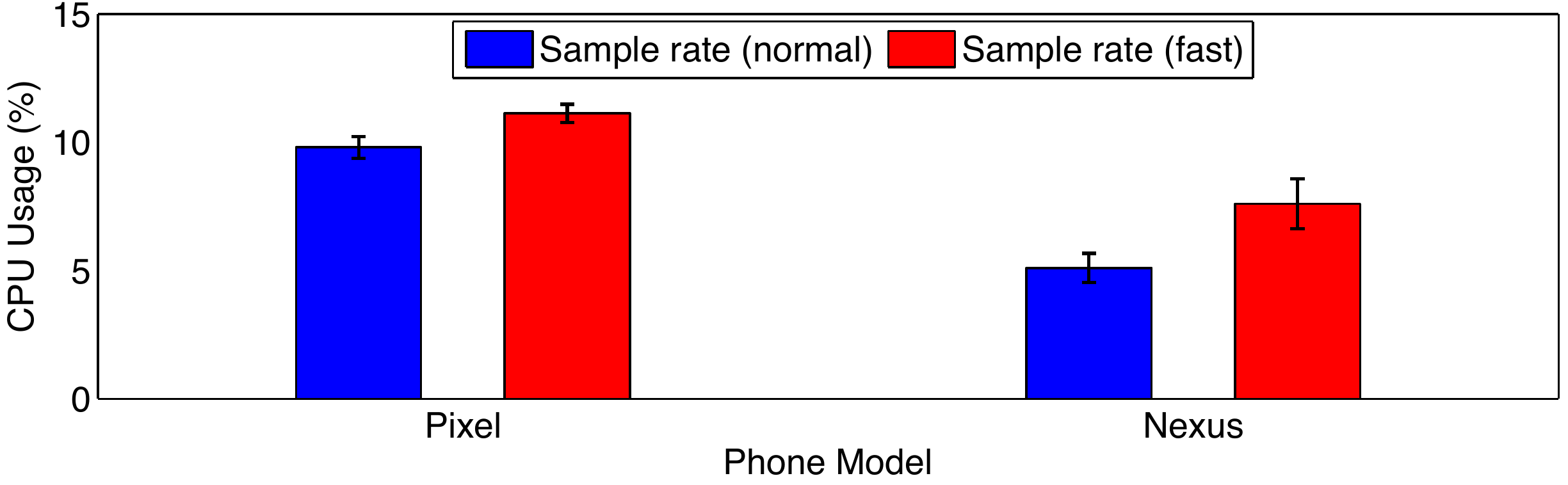}}
			\vspace{-1mm}
		\centerline{(a) CPU usage.}
	\end{minipage} 
\vskip 3pt
	\begin{minipage}[t]{0.50\textwidth}
		\centerline{\includegraphics[width=0.95\linewidth]{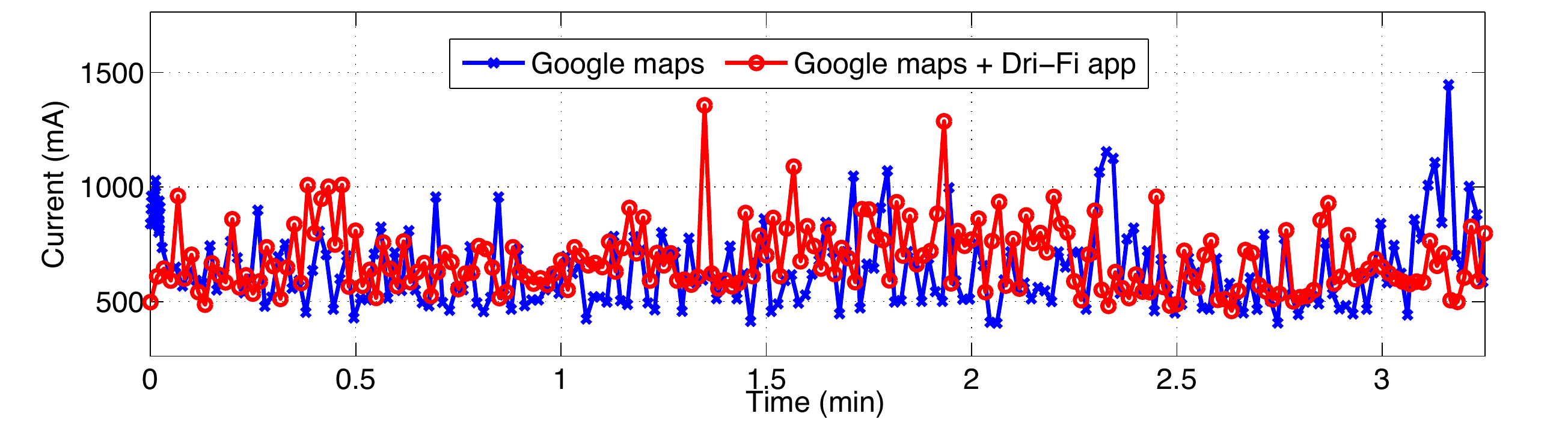}}
			\vspace{-1mm}
		\centerline{(b) Energy consumption.} 
	\end{minipage}
	\caption{Overhead of data collection app.}
	\label{fig:overhead_01}
	\vspace{-1em}
\end{figure}

To measure CPU usage, we recorded the CPU usage on both Google Pixel phone and 
Nexus 5X phone by using Android adb shell. 
To evaluate the extra overhead incurred by \name's data-collection module, 
which requires a bit higher sampling rate than usual, we compared the CPU usage 
of an application running with a normal IMU sampling rate (for detecting screen 
rotation) and with the sampling rate which {\name} uses: 100Hz. 
As shown in Fig.~\ref{fig:overhead_01}(a), albeit the increased sampling rate 
of \name, there were only small increases in the CPU usage; specifically, 2\% 
increase on a Pixel phone and 3.4\% increase on a Nexus 5X phone.
Since such an increased CPU usage was also occasionally observable even when 
running with a normal sampling rate, the increased CPU usage 
may not necessarily indicate (or let the victim know) that \name\ is running. 

We also examined the additionally consumed energy of using \name\ by measuring 
the current drawn in the smartphones. Fig.~\ref{fig:overhead_01}(b) shows the 
energy consumption on Pixel while \name\ was running in the background and 
utility applications (e.g. Google maps) were running in the foreground. 
Our results indicate that compared to the case where Google maps drew 767.10mA 
of current for navigation, \name\ drew 
only 49.60mA additional current. 
This 6.5\% extra energy consumption would be too minimal for the victim to 
notice.


Such small increases in CPU usage and energy consumption imply that if the 
compromised app/software originally has high overhead (e.g., navigation and 
social apps), then this marginal increase of these overhead caused by {\name} 
would be much less obvious. As a result, it will be even harder for the victim 
to notice such overheads. 

%% file: body/relatedwork.tex
\section{Related Work}\label{sec:relatedwork}

\subsection{Device/User Tracking on Mobile Devices}
While many researchers studied privacy invasion on mobile devices, their 
practicability in breaching the driver's privacy is limited by a common requirement: the mobile device's 
user/owner being tracked must also always be the driver.

A straightforward approach to identifying a mobile {\em device} is to 
track its unique identifier, such as IMEI, ICCID (SIM ID), etc., 
and/or cookies from its web browsers.
However, this approach is not practical since reading unique identifiers 
requires the user's permission and cookies can be easily deleted by the device user. 
To evade these, stealthier ways of tracking mobile devices have been explored. 
For example, Dey {\it et al.}~\cite{accel_NDSS2014} and Bojinov {\it et al.}~\cite{boneh_fingerprint} 
have shown adversaries to be able to track devices by utilizing the minor 
MEMS imperfection of the accelerometer. 
Resilient browser tracking approaches (e.g., evercookie~\cite{evercookie}) 
have also shown a practical way of tracking devices that are connected to the Internet. 

All of the above approaches are effective in identifying a device, but they share one common limitation 
in fingerprinting a driver: they cannot match the driving data with the user's identity 
when the user is riding as a passenger.
The motivating scenarios discussed in Section~\ref{subsec:motivating_cases} are 
exemplars that the device-oriented approaches fail to fingerprint drivers.
Note that even recent in-car phone localization methods --- e.g., determining 
whether the phone is placed in the driver's seat~\cite{phone_location} --- would 
also suffer from the same problem since the driver's phone can be placed 
anywhere in the car, e.g., inside a purse sitting in the passenger's seat.

Researchers have also been able to track the mobile user's identity based on his/her {\em behavioral} pattern(s). 
For example, Bo {\it et al.}~\cite{silent_sense} constructed a ``touch-based biometric'' 
based on the user's touch screen behavior and Herrmann {\it et al.}~\cite{dns_tracking} used the user's 
DNS traffic pattern to track him/her. 
Like device-oriented approaches, this type of tracking can also link the user's current behavior 
with his/her identity. Note, however, that the user interacting with the device being tracked 
(while driving) may not necessarily be the driver.

\subsection{Driver Fingerprinting Based on CAN Data}
Recently,  various in-vehicle sensor data havebeen used to fingerprint drivers 
\cite{yoshi2016, finger_sae17, single_turn2016, driver_inertial_sensor, driverID_gmm2007, knowmaster}.

Enev {\it et al.}~\cite{yoshi2016} investigated whether an adversary can identify/fingerprint the driver 
via in-vehicle (specifically CAN) data. By exploiting 18 or more types of CAN sensor data (e.g., brake pedal position, 
throttle position, engine speed) collected through the OBD-II port for at least 15 minutes, 
the adversary was shown to be able to fingerprint the driver with high accuracy. 
Similarly, the feasibility of fingerprinting the driver based on CAN data was shown in  \cite{finger_sae17}.

Kwak {\it et al.}~\cite{knowmaster} also exploited CAN data for driver
fingerprinting for an anti-theft purpose. In addition to the features proposed
in \cite{yoshi2016}, they exploited the mechanical features of
automotive parts (e.g., transmission oil temperature), thus 
enhancing the accuracy of driver fingerprinting. 

Hallac {\it et al.}~\cite{single_turn2016} exploited 12 different types
of CAN data for driver fingerprinting. Specifically, they proposed
a classification algorithm by exploiting simple to complex features such as
mean, standard deviation, and spectral components of the 12 different CAN sensor data.
That way, they were able to identify the driver with high accuracy within one turn, but {\em only 
when the number of different drivers were 2}.

Van Ly {\it et al.}~\cite{driver_inertial_sensor} also used in-car CAN data representing 
acceleration, brake, and turn signals to identify the driver. 
Other relevant work includes \cite{driverID_gmm2007} which  
used not only CAN data but also additional new features 
including the car-following distance and the sound information
when someone speaks inside the car for driver identification. 

These related studies have demonstrated the feasibility of driver 
fingerprinting, but {\em all} of them are based on in-vehicle data.
Both accessing and interpreting/decoding CAN data --- i.e., knowing which CAN 
ID contains which sensor values and how they are encoded --- are 
non-trivial problems.

In contrast, \name\ takes a very different approach from existing driver 
fingerprinting schemes by identifying the driver based solely on mobile IMUs. 
Since these are zero-permission sensors and available in all commodity 
smartphones, some off-the-shelf devices, and even in OBD-II dongles, from an 
adversary's point of view, the attack surface of fingerprinting drivers is much 
larger than that covered by existing schemes.
For example, any malicious app installed on the victim's smartphone 
can collect the IMU data.
Moreover,  \name\ can achieve
driver fingerprinting with high accuracy as soon as the driver makes 
a single turn. Even though \name\ can sometimes be 
incorrect after the first turn, as the driver makes more turns and thus
\name\ collects more data, its accuracy of fingerprinting improves significantly.

%% file: body/discussion.tex
\section{Discussion}
\label{sec:discussion}

\textbf{Number of drivers.}
We evaluated {\name} with up to 12 drivers.  
The fact that an adversary can accurately fingerprint the 
driver among such a number of candidates (even with access to only IMUs) 
implies a serious potential privacy risk. In most real-world scenarios, the 
maximum number of drivers for a given vehicle may not even be that large. 
Privately owned vehicles --- the most common scenario --- will most probably be 
driven by only a few people, such as family members.
Although the accuracy of Dri-Fi’s maneuver-based fingerprinting approach drops 
as the number of driver increases (Fig.~\ref{fig:maneuver}), \name\ can 
offset such a deterioration via a trip-based fingerprinting approach; as shown 
in Fig.~\ref{fig:tripbased}, \name\ can boost the accuracy 
with increasing number of turns.

\textbf{Countermeasures.}
To prevent an adversary from fingerprinting the driver via an IMU, 
one may add artificial noise to the sensor readings. 
Addition of noise does not necessarily have to be done continuously, but only 
when the driver is anticipated to start his turn. For example, as in 
Fig.~\ref{fig:sensor_turn}, when the absolute gyroscope 
readings exceed the threshold, $\delta_{bump}$, the device can be configured to 
add noise. Accordingly, an adversary exploiting \name\ would be unable to 
extract accurate measurements from a vehicle turn and thus fail in driver 
fingerprinting. For smartphones, such an approach should be implemented in the 
OS-level, if there are no other apps using IMU sensors for ``good 
purposes'' while driving. 
Another countermeasure (in case of a smartphone) is to request permission for use of IMU sensors 
when installing the app, as discussed in \cite{inferroute}.

\textbf{Limitations.}
With the coordinate alignment discussed in Sec.~\ref{subsec:data_collection}, 
{\name} can analyze the data from a consistent viewpoint, i.e., according to 
the geo-frame coordinate, no matter what the posture of the device has.
However, if the device always moves during driving (e.g., the driver's 
smartwatch), it may introduce noise in the measurements, thus rendering \name\ 
to be less accurate. 
Moreover, depending on the driver's emotion, sense of urgency, and physical 
status, \name's accuracy may vary as well.
In fact, the false-positives/negatives in \name's fingerprinting might  
occur for these reasons. So, in future we would like to 
conduct a detailed study of the effects of such factors on \name's accuracy.
%
%

%% file: body/conclusion.tex
\section{Conclusion}
\label{sec:conclusion}
In this paper, we presented \name, a driving data analytic engine that an 
adversary can exploit to fingerprint the driver within only one vehicle turn, 
and most importantly 
using only zero-permission mobile IMUs.
\name\ achieves this by capturing new representative features of a driver's unique way of making turns.
Via extensive evaluations, \name's extracted features 
are shown to represent the driver's unique turning style and thus function as the key in fingerprinting the driver.
Such a feature of \name\ implies a significant expansion of the attack vector; 
an adversary can identify the driver with access to not only in-car data but 
also to IMU-equipped devices. More importantly, it implies a practical, 
serious, and a yet uncovered privacy threat. 
Accordingly, we suggest both academia and industry to 
be wary of such a threat and thus make concerted efforts to develop 
countermeasures.

%% file: body/reference.tex
\bibliography{ieee-reference}